\documentclass[12pt, draftclsnofoot, onecolumn]{IEEEtran}

\usepackage{cite}
\usepackage{graphicx}
\usepackage{graphicx}
\usepackage{float}
\usepackage{algorithm}
\usepackage{algorithmic}
\usepackage{flushend}
\usepackage{mathrsfs}
\usepackage{amsfonts}
\usepackage{array}
\usepackage{amssymb,amsmath}
\usepackage{longtable}
\usepackage{rotating}
\usepackage{multirow}
\usepackage{cite}
\usepackage{float}
\usepackage[marginal]{footmisc}
\usepackage{stfloats}
\usepackage{setspace}
\usepackage{cases}
\usepackage{color}
\usepackage{booktabs}
\usepackage{threeparttable}

\def\BibTeX{{\rm B\kern-.05em{\sc i\kern-.025em b}\kern-.08em
    T\kern-.1667em\lower.7ex\hbox{E}\kern-.125emX}}

\begin{document}
%
\title{Machine Learning-based Signal Detection for PMH Signals in Load-modulated MIMO System}


\author{\IEEEauthorblockN{Jinle Zhu, Qiang Li, Li Hu, \\
Hongyang Chen, \textit{Senior Member}, \textit{IEEE}, and Nirwan Ansari, \textit{Fellow}, \textit{IEEE}}}

\maketitle
\vspace{-6em}

\begin{abstract}
Phase Modulation on the Hypersphere (PMH) is a power efficient modulation scheme for the \textit{load-modulated}  multiple-input multiple-output (MIMO) transmitters with central power amplifiers (CPA).
However, it is difficult to obtain the precise channel state
information (CSI), and the traditional optimal maximum likelihood
(ML) detection scheme incurs high complexity which
increases exponentially with the number of antennas and the
number of bits carried per antenna in the PMH modulation.
To detect the PMH signals without knowing the prior CSI, we first propose a signal detection scheme, termed as the hypersphere clustering scheme based on the expectation maximization (EM) algorithm with maximum likelihood detection (HEM-ML). By leveraging machine learning, the proposed detection scheme can accurately obtain information of the channel from a few of the received symbols with little resource cost and achieve comparable detection results as that of the optimal ML detector. To further reduce the computational complexity in the ML detection in HEM-ML, we also propose the second signal detection scheme, termed as the hypersphere clustering scheme based on the EM algorithm with KD-tree detection (HEM-KD). The CSI obtained from the EM algorithm is used to build a spatial KD-tree receiver codebook and the signal detection problem can be transformed into a nearest neighbor search (NNS) problem. The detection complexity of HEM-KD is significantly reduced without any detection performance loss as compared to HEM-ML. Extensive simulation results verify the effectiveness of our proposed detection schemes.
\end{abstract}
\vspace{-1.5em}

\vspace{-1em}

\section{Introduction}
The fifth generation (5G) wireless communication network is forecasted to provide over 1000 times higher capacity than the current system. In addition to dramatically expanding the available bandwidth, multiple-input multiple-output (MIMO) technology is playing a key role in improving the spectral efficiency (SE) and enhancing the throughput in the future wireless cellular communication systems \cite{5G}. 
\footnote{
This work was supported in part by National Natural Science Foundation of China(No.61571082) \& National Key R\&D Program of China(No.254).}

This ambitious goal will however cause an inevitable energy consumption problem, thus limiting the number of the antennas at the base station (BS) and the user terminals in practice \cite{power}. In the traditional design of the MIMO transceivers, each antenna is connected with one distinct radio frequency (RF) chain which includes a power amplifier (PA). This kind of structure enables the power consumption of the transmission to grow linearly with the number of the antennas. In addition, the use of Orthogonal Frequency Division Multiplexing (OFDM) signals in massive MIMO systems leads to a high peak-to-average power ratios (PAPR) and exacerbates the costs of PAs, thus reducing the power efficiency. On the other hand, to alleviate the effects of mutual coupling and correlated fading, the antennas should be set at least half of a wavelength apart from each other, which will inevitably cause the size problem \cite{antenna}.

Sedaghat \textit{et al.} \cite{singleRF} discussed a novel single-RF transmitter structure called \textit{load-modulated} MIMO transmitter to solve the aforementioned issues. In contrast with the ESPAR (Electronically Steerable Parasitic Array Radiator) transmitter proposed in \cite{ESPAR}, the transmitter design in \cite{singleRF} can support any type of modulation and allow softly changing the currents on antennas. Furthermore, the central power amplifier (CPA) generates a fixed amount of instantaneous total power which enables the transmitter to work with a high efficiency. Afterwords, they \cite{MCPM}\cite{PMH} derived a novel modulation scheme called phase modulation on the hypersphere (PMH) for the single-RF MIMO transmitter in \cite{singleRF}. Distributing input signal with PMH on a hypersphere can achieve the close channel capacity in terms of the Gaussian input \cite{capacity}. The number of the transmitter codewords is exponential with the number of antennas and the number of bits carried per antenna, thus enabling the PMH symbols to achieve a high spectral efficiency. References \cite{MCPM} and \cite{PMH} mainly focused on the information capacity study and the pulse shaping methods of the PMH modulation scheme at the transmitter. The authors straightforwardly used the optimal maximum likelihood (ML) detector at the receiver.

While the ML detector can achieve the optimal bit error rate (BER), the detection process requires the prior channel information which is difficult to acquire. Using pilot symbols can help estimate the channel information, but it will lead to an unavoidable waste of power and resource. In addition, in the calculation of the Euclidean distance in the ML detection process, the number of the multiplication operations increases exponentially with the number of antennas and the number of bits carried per antenna; this is computationally expensive, thus compromising efficiency of the PMH modulation. The drawback of this modulation is exacerbated for a MIMO system with a large number of antennas or high order modulation.

{From the up-to-date research, instead of acquiring the channel information aided by sequence datas or detecting the signals in a traditional complicated fashion, machine learning \cite{MLL} can be leveraged by performing a signal-driven approach to extract desired features from the known received signals \cite{ML1,ML2}. Inspired by some successful applications of machine learning in wireless communications \cite{ML3}, we propose to employ machine learning to handle the detection problem of PMH signals.}
The fact that the PMH symbols are distributed on a multidimensional hypersphere {falling into several clusters} has inspired us to utilize the clustering algorithms {in machine learning} to detect the symbols. {The clustering algorithm can learn the identification information carried by the received signals with the spatial clustering features}. The received symbols follow a Gaussian mixture model (GMM) in which the feature extraction problem can be solved based on the expectation maximization (EM) algorithm \cite{PML}\cite{PC}.

We propose two schemes to detect the PMH symbols. One is the hypersphere clustering scheme based on the EM algorithm with ML detection (HEM-ML). First, few pilot sequences are used for the initialization in the EM algorithm. Then, the EM algorithm is implemented to jointly update the posterior probability of the received signals and refine the desired channel information. Besides, only a small number of the received symbols in the EM algorithm can extract the channel feature well. Thus, to reduce the computation operations in the EM algorithm, we use a few of the received symbols to engage in the EM algorithm. The remaining received symbols can be detected by the traditional ML detection with the estimated channel information obtained from the EM algorithm.

Though we solve the problem of the unknown channel state information (CSI) in HEM-ML, most of the received signals in the frame are still detected in a brute-force way. To further reduce the detection complexity, we propose the second algorithm, termed as the hypersphere clustering scheme based on the EM algorithm with KD-tree detection (HEM-KD). The channel information obtained by the EM algorithm is used to construct the spatial KD-tree codebook to store the receiver codewords. The multidimensional feature of the PMH symbols inspires us to formulate the detection process as a problem of finding the nearest point to the received signal point in the multidimensional tree. Thus, the signal detection problem is transformed into a nearest neighbor search (NNS) problem.
Santipach and Mamat \cite{TS} constructed a quantized vector codebook by using a KD-tree in a CDMA system. The receiver chooses the best leaf node as a feedback and sends it to the transmitter. The KD-tree in \cite{TS} is an unbalanced tree and only the leaf nodes are taken into consideration at the receiver.
In contrast from \cite{TS}, the spacial nature of the transmitted PMH signals are fully utilized for constructing a balanced spatial KD-tree codebook in our paper. The element of each antenna is a coordinate to a PMH signal, and thus a PMH signal can be stored in a spatial KD-tree according to the multidimensional coordinates.

The EM algorithm, the KD-tree and the NNS problem have been widely employed in the field of image processing and computer graphics. In this paper, these machine learning techniques are utilized to introduce a new perspective in signal detection. The main contributions of this paper are summarized as follows:
\begin{itemize}
\item {We propose an approach to detect PMH signals without knowing the prior CSI. The feature of the constellation points distributed on a hypersphere allows us to incorporate the clustering algorithm (EM algorithm) into the detection schemes HEM-ML and HEM-KD with low initialization and iteration overhead. The clustering method is able to fully exploit the channel information carried by a small number of received symbols to achieve a precise channel estimation with high resource efficiency.
\item Second, for multidimensional symbol detection, we propose a systematic detection process based on KD-tree and NNS. A novel spatial codebook design based on KD-tree for the PMH signals is proposed. The balanced KD-tree stores constellation points of the receiver codewords in the multidimensional space. Then, the detection process is to simply search for the nearest neighbor point by traversing down the tree according to the dimension indexes, thus resulting in linear detection complexity without any loss of detection precision as compared with a traditional ML search method.
\item Finally, extensive numerical results are presented to verify the effectiveness of the proposed detection schemes HEM-ML and HEM-KD. The proposed schemes achieve bit-error-rate (BER) performances close to the optimal ML detector especially in high SNR case. The detection complexity of the proposed schemes is also analyzed.}
\end{itemize}

The rest of this paper is organized as follows. Section \uppercase\expandafter{\romannumeral2} introduces the \textit{load-modulated} system model, the PMH modulation and the method to obtain the PMH transmitted codewords. Section \uppercase\expandafter{\romannumeral3} elaborates on the probability model of the received symbols and presents the optimal ML detector with perfect CSI, the LS detector with few pilot sequences and the proposed HEM-ML signal detection scheme. Section \uppercase\expandafter{\romannumeral4} describes the proposed HEM-KD signal detection scheme to further reduce the detection complexity. Section \uppercase\expandafter{\romannumeral5} provides the computational complexity comparisons of different signal detection schemes. Section \uppercase\expandafter{\romannumeral6} shows the simulation results to verify our proposed schemes and the conclusion is made in Section \uppercase\expandafter{\romannumeral7}.

Notation: We use upper-case boldface letters, lower-case boldface letters and lower-case $\textbf{X}$, $\textbf{x}$ and $x$ to denote a matrix, vector and scalar, respectively; $\mathbb{C}^{i\times j}$ denotes the space of ${i\times j}$ complex matrix. $(\cdot)^H$, $(\cdot)^{-1}$ and $\|\cdot\|$ denote the conjugate transpose, matrix inversion and Frobenius norm operation, respectively. $\textbf{I}_{M}$ is the $M\times M$ identity matrix.

\section{System model}

We consider a \textit{load-modulated} single-RF MIMO transmitter \cite{singleRF} as shown in Fig. 1 and the downlink of a point-to-point (P2P) communication system. The transmitter is equipped with $N$ antenna elements fed by a single common power amplifier via $N$ load modulators ${\rm{M}_{{\textit{n}}}}$ (${{n}} = 1,...,N$). 
The data is carried by the signals using PMH modulation which is a generalization of phase modulation from the complex unit circle to the hypersphere \cite{MCPM}. The squared Euclidean norm of the transmitted signals is fixed to the constant sum power such that the peak to average ratio of the sum power (PASPR) remains low even to 0dB.

\begin{figure}[!htb]
\renewcommand{\figurename}{Fig.}
\centering
\includegraphics[width=7.5cm]{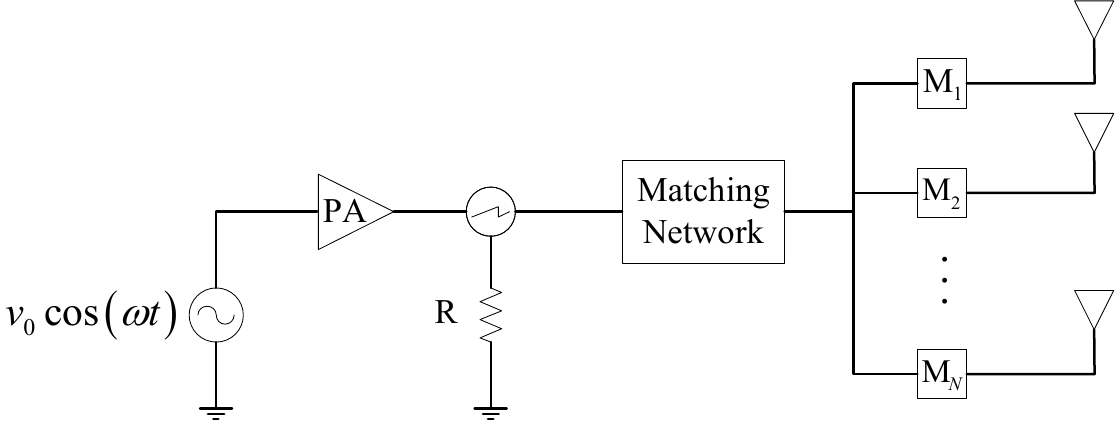}%
\label{receiver}
\vspace{-1em}
\caption{The MIMO transmitter with single-RF chain proposed in \cite{singleRF}.}
\end{figure}
\vspace{-1em}

Without loss in generality, we assume the transmitted signals satisfy ${\left\| {\textbf{\textit{x}}}_t \right\|^2}=N$, where $\textbf{\textit{x}}_t \in {\cal S}$ is a $\sqrt N$-power $K$-ary transmitted PMH symbol in $N \times 1$ dimension at discrete time $t$ and the \textit{n}-th element of $\textbf{\textit{x}}_t$ denotes the signal of the \textit{n}-th antenna $(n=1,2,...,N)$.
${\cal S}=\{{\textbf{\textit{s}}_1},{\textbf{\textit{s}}_2},...,{\textbf{\textit{s}}_K}\}$  is the codeword alphabet, i.e., the transmitter codebook, which is known by both the transmitter and the receiver. The codewords ${\textbf{\textit{s}}_i}=[{{{s}}_i^1,{{s}}_i^2,...,{{s}}_i^{N}}]^T$ for $i = {1,2,...,K}$ are the transmitter symbol with PMH modulation in $N \times 1$ dimension. It has been proven that the maximum capacity is achieved when the transmitter signal $\textbf{\textit{s}}_i$ is uniformly distributed on the surface of a hypersphere
${\mho_N}\left( N \right)$ where ${\mho_N}\left( N \right){\rm{ = }}\left\{ {\textbf{\textit{s}} \in {C^N}\left| {{{\left\| \textbf{\textit{s}} \right\|}^2} = N} \right.} \right\}$ denotes the hypersphere of radius $\sqrt N$ in $N$ complex dimensions \cite{MCPM}.
From the perspective of regarding the real part and imaginary part of the codeword entries in different dimensions, the transmitter signal constellation is in a 2\textit{N}-dimensional real-valued spherical coordinate system.
The received signal $\textbf{\textit{y}}_t$ at discrete time $t$ can be presented as
\begin{equation}\label{system model}
\textbf{\textit{y}}_t = \textbf{\textit{H}}\textbf{\textit{x}}_t + \textbf{\textit{n}}_t,
\end{equation}
where $\textbf{\textit{H}}$ is the $M \times N$ channel matrix and $\textbf{\textit{n}}_t \in {\mathbb C^{M \times 1}}$ is the \textit{i.i.d.} (independent and identically distributed) white Gaussian noise vector at discrete time $t$ with variance ${\bf{\Sigma} }={\sigma ^2}\textbf{I}_M$. {\textit{M} is the number of antennas of the receiver. If $M<N$, some information of the transmitted signals is lost at the receiver due to reduction of the dimension that will result in performance degradation, and thus we assume $M\geqslant N$.}

In this paper, we assume that the channel is quasi-static, which means the channel stays the same in one frame with $T$ symbols. Then, the transmitted symbol matrix of a frame is $\textbf{\textit{X}}=[\textbf{\textit{x}}_1,\textbf{\textit{x}}_2,...,$
$\textbf{\textit{x}}_t]$ and the received symbol matrix is $\textbf{\textit{Y}}=[\textbf{\textit{y}}_1,\textbf{\textit{y}}_2,...,\textbf{\textit{y}}_t]$.
There are $K{\rm{ = }}{2^{{{k}}N}}$ codewords in the alphabet ${\cal S}$ where \textit{k} is the number of bit transmitted by each antenna in each PMH symbol. Thus, the transmitter sends ${kN}$ bits in each symbol time.
The size of the codebook grows exponentially with the number of the antennas and the number of bit carried by each antenna.
The transmitter codewords $\textbf{\textit{s}}_i$, uniformly distributed on the surface of a hypersphere ${\mho_N}\left( N \right)$, can be encoded into the spherical codes $C(K,2N,{\theta _{\min }})$, i.e., $\textit{K}$ code vectors in $\mathbb{R}^{2N}$ with at least an angular separation of ${\theta _{\min }}$.
The spherical code words with the maximum angular separation have the best error exponent\cite{sphere}.
However, the exact solution to the spherical codes with the best error exponent can only be solved for a few dimensions\cite{SC}.

Sedaghat \textit{et al.} \cite{PMH} proposed two sub-optimal methods to construct spherical codes. The first method is to generate a large number of uniformly distributed points on the hypersphere and then to cluster them by the spherical K-means algorithm into \textit{K} points on the hypersphere as the transmitter codewords in the alphabet. The spherical codes generated by this method are called the spherical K-means codes. The second method is the Equal-Area (EQ) sphere partition algorithm taken from \cite{EQ}. The simulation results in \cite{PMH}  proved that the codewords obtained by the spherical K-means algorithm have much better performance than the EQ codewords. Thus, in this paper, we adopt the first method, namely, the spherical K-means algorithm, to generate spherical codes. As noted in \cite{PMH}, Gray labels mapping derived by the algorithm proposed in \cite{gray} is also adopted in this paper for better error performance.

The detection scheme is critical to ensure high quality PMH modulation with high spectral efficiency. The authors in \cite{MCPM,PMH} analyzed the capacity and the shaping pulse problem of the PMH signals and straightforwardly adopted the optimal ML detection scheme at the receiver assuming the CSI is perfectly known. However, it is difficult to acquire the CSI of a large number antenna MIMO system. Besides, the computational complexity of the detection process in the ML detection of the PMH signals rapidly increases with the number of antennas. Furthermore, the size of the transmitter codebook of the PMH signal also increases exponentially with the number of bits per antenna. It is thus essential to design an efficient signal detection scheme for detecting the PMH signal without the knowledge of channel.

In the next sections, we will introduce our proposed machine learning-based detection schemes which are able to extract the channel information accurately from the received symbols and detect the received signals by exploiting the multi-dimensional spatial structure of PMH symbols.

\section{Hypersphere clustering scheme based on EM Algorithm with ML detection (HEM-ML)}

As introduced in Section \uppercase\expandafter{\romannumeral2}, the constellation points of PMH signals are uniformly distributed on an $N-$dimensional hypersphere. We assume that the transmitter codewords are equiprobable. Inspired by the spatial distribution of the PMH signals, we use the EM algorithm to cluster the received symbols, i.e., to cluster the points on the hypersphere. To obtain the channel parameter, a traditional method is the least square (LS) estimation which utilizes the pilot sequence. When compared to the LS algorithm, the EM estimation is able to better exploit the information carried by the received signals that can lead to either better channel parameter estimation or less number of pilot sequences. {We will first introduce the traditional optimal ML detector with perfect CSI and the LS detector as performance benchmarks and then describe our proposed HEM-ML detection scheme.}
\vspace{-1em}
\subsection{Optimal ML detection scheme with perfect CSI}

In this subsection, we briefly introduce the optimal ML detector. When the prior probabilities of each codeword are equal, the ML detector can achieve the best error rate performance \cite{ML}. The ML detector searches the optimal signal in the codebook in a brute force method by directly comparing the ML metrics.

As shown in (\ref{system model}), in a frame, the received symbol at discrete time $t$ follows a complex Gaussian distribution with the mean of $\textbf{\textit{H}}\textbf{\textit{x}}_t$ and the variance of ${\sigma ^2}\textbf{I}_M$, i.e., $\textbf{\textit{y}}_t\sim \mathcal{CN}(\textbf{\textit{H}}\textbf{\textit{x}}_t,{\bf{\Sigma} })$ where
$\mathcal{CN}\left( {{\textbf{\textit{y}}_t}|\textbf{\textit{H}}{\textbf{\textit{x}}_t},{\bf\Sigma }} \right)
{\rm{ = }}
\frac{1}{{{\pi ^M}\left| {\bf\Sigma} \right|}}\exp \left\{ {\! - {{\left( {{\textbf{\textit{y}}_t} - \textbf{H}{\textbf{\textit{x}}_t}} \right)}^H}\!{\bf\Sigma ^{ - 1}}\!\left( {{\textbf{\textit{y}}_t} - \textbf{\textit{H}}{\textbf{\textit{x}}_t}} \right)}\! \right\}.$
With the perfect CSI, the decision rule of the optimal ML detector is given by
\begin{equation}\label{ML}
{\hat {\textbf{\textit{x}}}_{{\rm ML},t}} = \mathop {\arg \max }\limits_{{\textbf{\textit{s}}_i} \in {\cal S}} \mathcal{CN}\left( {{\textbf{\textit{y}}_t}|\textbf{\textit{H}}{\textbf{\textit{s}}_i},{\bf\Sigma }} \right).
\end{equation}

{However, the difficulty of acquiring the CSI of the system renders the optimal ML detector practically infeasible.}

\subsection{Least Square Algorithm with Pilots Sequence}
When it is difficult to obtain the CSI, there is a known signal detection technique based on the LS estimation. The pilot sequence is essential to acquire the CSI at the expense of bandwidth and power especially when the length of the required pilot sequence is large.

The optimal transmitted pilot symbol matrix is $\textit{\textbf{X}}_{pi}=[\textit{\textbf{x}}_{pi,1},\textit{\textbf{x}}_{pi,2},...,\textit{\textbf{x}}_{pi,L}]\in \mathbb{C}^{N\times L}$ which satifies $\textit{\textbf{X}}_{pi}\textit{\textbf{X}}_{pi}^{H}=\textit{\textbf{I}}_N$ where $L$ is the length of the pilot sequence and $\textit{\textbf{x}}_{pi,l}$ $(l=1,2,...,L)$ is the $l$-th transmitted pilot symbol. The corresponding received symbol $\textit{\textbf{y}}_{pi,l}$ is given by $\textit{\textbf{y}}_{pi,l}=\textit{\textbf{H}}\textit{\textbf{x}}_{pi,l}+\textit{\textbf{n}}_{pi,l}.$
With all the $L$ pilot symbols, the received pilot symbol matrix is $\textit{\textbf{Y}}_{pi}=[\textit{\textbf{y}}_{pi,1},\textit{\textbf{y}}_{pi,2},...,\textit{\textbf{y}}_{pi,L}]$. Then, the estimated channel matrix $\textit{\textbf{H}}_{\rm{LS}}$ is given by
\begin{equation}\label{HLS}
\textit{\textbf{H}}_{{\rm LS}} = \textit{\textbf{Y}}_{pi}(\textit{\textbf{X}}_{pi}^{H}\textit{\textbf{X}}_{pi})^{-1}\textit{\textbf{X}}^{H}_{pi}.
\end{equation}

The noise variance can be estimated according to the ML criterion after the estimation of the channel matrix $\textit{\textbf{H}}_{{\rm LS}}$. The estimated noise variance ${\bf\Sigma}_{\rm LS}$ is given by
\begin{equation}\label{SLS}
{\bf\Sigma}_{\rm LS} = \frac{1}{L}\sum\limits_{{l} = 1}^L\\
{[(\textit{\textbf{y}}_{pi,l}-\textit{\textbf{H}}_{{\rm LS}}\textit{\textbf{x}}_{{pi,l}})(\textit{\textbf{y}}_{pi,l}-\textit{\textbf{H}}_{{\rm LS}}\textit{\textbf{x}}_{{pi,l}})^{H}]}.
\end{equation}

With the estimated CSI, the decision rule for ${\textbf{\textit{y}}_t}$ based on the LS estimation is given by
\begin{equation}\label{LS}
{\hat {\textbf{\textit{x}}}_{{\rm LS},t}} = \mathop {\arg \max }\limits_{{\textbf{\textit{s}}_i} \in {\cal S}} \mathcal{CN}\left( {{\textbf{\textit{y}}_t}|\textbf{\textit{H}}_{{\rm LS}}{\textbf{\textit{s}}_i},{\bf\Sigma }_{\rm LS}} \right).
\end{equation}

The performance of the detector based on the LS estimator depends on the number of the pilot sequences, which will incur the pilot overhead problem if a high quality estimated channel is needed.

\subsection{EM-based Signal Detection with few Pilots}

{When the CSI is unavailable, the infeasibility of the optimal ML detector and the low efficiency in communication resource utilization of the LS detector motivate us to design a new signal detection scheme for the PMH signal.} By noticing that the constellation points of the codewords naturally fall into clusters, we utilize clustering algorithms in machine learning to cluster the received signals of the receiver.
In this subsection, we introduce our proposed HEM-ML detector. We first assume that all received symbols in a frame are used in the EM algorithm to jointly estimate the channel matrix and cluster the received symbols.

\subsubsection{Gaussian Mixture Model (GMM)}
At the receiver, the received symbols in a frame follow a $K$-component complex Gaussian mixture distribution. The probability density function (PDF) of ${\textbf{\textit{y}}_t}$ can be represented as a GMM, i.e.,
\begin{align}\label{GMM}
p\left( {{\textbf{\textit{y}}_t}|\textbf{\textit{H}}\textbf{\textit{x}}_t,{\bf\Sigma }} \right){\rm{ = }}\sum\limits_{{i} = 1}^K \rho_i\mathcal{CN}\left( {{\textbf{\textit{y}}_t}|\textbf{\textit{H}}{\textbf{\textit{s}}_{i}},{\bf\Sigma }} \right),
\end{align}
where $\rho_i$ is the probability that the transmitter sends out $\textbf{\textit{s}}_i$. Since the transmitter codewords are equiprobable, we have $\rho_i=\frac{1}{K}$.

\subsubsection{Joint GMM-based EM channel estimation and symbol detection}
We now show the details of the proposed HEM-ML scheme. The EM algorithm is used to extract the channel information carried by the received symbols. With the PDF of ${\textbf{\textit{y}}_t}$ expressed in (\ref{GMM}), the PDF of the symbols in a frame is
$p\left( {{\textbf{\textit{Y}}}|\textbf{\textit{H}}\textbf{\textit{X}},{\bf\Sigma }} \right){\rm{ = }}\prod\limits_{t = 1}^T {\sum\limits_{{i} = 1}^K \rho_i\mathcal{CN}\left( {{\textbf{\textit{y}}_t}|\textbf{\textit{H}}{\textbf{\textit{s}}_{i}},{\bf\Sigma }} \right)}.$

The parameters we need to estimate are the channel matrix $\textbf{\textit{H}}$ and the variance of the noise ${\bf\Sigma }$ which are denoted as $\theta = \{\textit{\textbf{H}},\bf\Sigma\}$. The log-likelihood function of $\theta$ is given by
\begin{align}\label{log}
LL\left( {\theta |\textbf{\textit{Y}}} \right) = \sum\limits_{t = 1}^T {\ln \sum\limits_{i = 1}^K {{\rho _i}\mathcal{CN}\left( {{\textit{\textbf{y}}_t}|\textbf{\textit{H}}{\textbf{\textit{s}}_{i}},\bf\Sigma } \right)} }.
\end{align}

By maximizing (\ref{log}), we can obtain the estimated parameters. However, the log-likelihood function is non-convex. The lack of the knowledge of the transmitted symbol does exacerbate the ML estimation. We resort to the EM algorithm which is an iterative method that yields a suboptimal solution.

In the EM algorithm, the received symbols $\textbf{\textit{Y}}$ are incomplete data because the corresponding transmitted signals $\textbf{\textit{X}}$ are unknown. We introduce a hidden variable $\textbf{\textit{Z}}{\rm{ = }}\left[ {{\textit{\textbf{z}}_1},{\textit{\textbf{z}}_2},...,{\textit{\textbf{z}}_T}} \right]$ where ${\textbf{\textit{z}}_t} = {\left[ {{z_{t,1}},{z_{t,2}},...,{z_{t,K}}} \right]^T}$. For the $t$-th transmitted symbol ${\textit{\textbf{x}}_t}$, we have ${{z_{t,i}}=1}$ for $\textbf{\textit{x}}_t = \textbf{\textit{s}}_i$ and ${{z_{t,i}}=0}$ for otherwise situation. The received symbols $\textbf{\textit{Y}}$ together with the hidden variable $\textbf{\textit{Z}}$ form the complete data set to facilitate the calculation of the parameters in (\ref{log}). In every symbol period, the probability of sending $\textbf{\textit{s}}_k$ at the transmitter is $p\left( {{z_{t,i}} = 1} \right) = {\rho _i} = \frac{1}{K}$.
The PDF of the complete data set is given by
$\nonumber p\left( {\textbf{\textit{Y}},\textbf{\textit{Z}}|\theta } \right) = \prod\limits_{t = 1}^T {\left( {\prod\limits_{i = 1}^K {{{\left[ {{\rho _i}\mathcal{CN}\left( {{\textit{\textbf{y}}_t}|\textit{\textbf{H}}{\textit{\textbf{s}}_i},\bf\Sigma } \right)} \right]}^{{z_{t,i}}}}} } \right)}$
and the log-likelihood function of the complete data set is written by
\begin{align}\label{likeli}
LL \left( {\theta | \textbf{\textit{Y}},\textbf{\textit{Z}}} \right) \!= \!\sum\limits_{t = 1}^T {\sum\limits_{i = 1}^K {{z_{t,i}}\!\left( {\ln {\rho _i} \!+\! \ln \mathcal{CN}\left( {{\textit{\textbf{y}}_t}|\textit{\textbf{H}}{\textit{\textbf{s}}_i},\bf\Sigma } \right)} \right)} }.
\end{align}

The EM algorithm provides a framework to iteratively update the probabilities on the symbols sent by the transmitter and estimate the parameter $\theta$ by carrying out the expectation step (E-step) and the maximization step (M-step) until the log-likelihood function (\ref{likeli}) or the parameter $\theta$ converges. The specific steps are described as follows where the superscript/subscript ($ite$) refers to the $ite$-th iteration.
\begin{itemize}
  \item E-step: At the E-step, we obtain the expectation of the hidden parameter $\textbf{\textit{Z}}$ as
      \begin{align}\label{E}
      & Q({\theta ^{\left( ite+1 \right)}}\mid{\theta ^{\left( ite \right)}})
      ={\mathbb{E}_{\textbf{\textit{Z}}{\rm{|}}\textbf{\textit{Y}},{\theta ^{\left( ite \right)}}}}\left[ { LL\left( {\textit{\textbf{Y}},\textit{\textbf{Z}}|{\theta ^{\left( ite+1 \right)}}} \right)} \right]\\
      = &\nonumber\sum\limits_{t = 1}^T {\sum\limits_{i = 1}^K {{\tau^{(ite)}_{t,i}}\left( {\ln {\rho_i} + \ln \mathcal{CN}\left( {{\textit{\textbf{y}}_t}|\textit{\textbf{H}}^{\left( ite+1 \right)}{\textit{\textbf{s}}_i},{\bf{\Sigma}}^{\left( ite+1 \right)} } \right)} \right)} }
      \end{align}
      where
      \begin{align}\label{tau}
      \tau^{(ite)}_{t,i}
      =\nonumber &\mathbb{E}\left[ {{z_{t,i}}|\textit{\textbf{Y}},{\theta ^{\left( ite \right)}}} \right]
      =p\left( {{z_{t,i}} = 1|\textit{\textbf{Y}},{\theta ^{\left( ite \right)}}} \right)
      =\frac{{p\left( {{z_{t,i}} = 1,\textit{\textbf{Y}},{\theta ^{\left( ite \right)}}} \right)}}{{p\left( {\textit{\textbf{Y}},{\theta ^{\left( ite \right)}}} \right)}}\\
      = &\frac{{p\left( {{z_{t,i}} = 1,\textit{\textbf{Y}},{\theta ^{\left( ite \right)}}} \right)}}{{\sum\limits_{q = 1}^K {p\left( {{z_{t,q}} = 1,\textit{\textbf{Y}},{\theta ^{\left( ite \right)}}} \right)} }}
      =\frac{{\mathcal{CN}\left( {{\textit{\textbf{y}}_t}|{\textit{\textbf{H}}^{\left( ite \right)}}{\textit{\textbf{s}}_i},{\bf\Sigma}^{(ite)} } \right)}}{{\sum\limits_{q = 1}^K {\mathcal{CN}\left( {{\textit{\textbf{y}}_t}|{\textit{\textbf{H}}^{\left( ite \right)}}{\textit{\textbf{s}}_q},{\bf\Sigma}^{(ite)} } \right)} }}.
      \end{align}

  \item M-step: To obtain the updated parameter $\theta^{(ite+1)}=\{\textit{\textbf{H}}^{(ite+1)},{\bf\Sigma}^{(ite+1)}\}$, we take the first-order derivatives of $Q({\theta ^{\left( ite+1 \right)}}\mid{\theta ^{\left( ite \right)}})$ obtained in E-step for $\textit{\textbf{H}}^{(ite+1)}$ and ${\sigma}^2_{(ite+1)}$ as
      \begin{equation}\label{H}
      \textit{\textbf{H}}^{\left( {ite{\rm{ + }}1} \right)}{\rm{ = }}\frac{{\sum\limits_{t = 1}^T {\sum\limits_{i = 1}^K {{\tau _{t,i}}{\textit{\textbf{y}}_t}\textit{\textbf{s}}_i^H} } }}{{\sum\limits_{t = 1}^T {\sum\limits_{i = 1}^K {{\tau _{t,i}}{\textit{\textbf{s}}_i}\textit{\textbf{s}}_i^H} } }},
      \end{equation}
      \begin{equation}\label{sigma}
      \sigma _{{\left( {ite + 1} \right)}}^2{\rm{ = }}\frac{{\sum\limits_{t = 1}^T {\sum\limits_{i = 1}^K {{\tau _{t,i}}\left\| {{\textit{\textbf{y}}_t} - {\textit{\textbf{H}}^{\left( {ite{\rm{ + }}1} \right)}}{\textit{\textbf{s}}_i}} \right\|_2^2} } }}{M{\sum\limits_{t = 1}^T {\sum\limits_{i = 1}^K {{\tau _{t,i}}} } }}.
      \end{equation}
\end{itemize}
Note that we take the derivative of $\sigma^{2}_{(ite+1)}$ instead of ${\bf\Sigma}^{(ite+1)}$ straightly because we assume that the variance of the noise of each antenna at the receiver is the same. We take the derivatives of $\sigma^2$ to utilize this implied condition to ensure convergence of the variance to the right stationary point. Then, the variance matrix of the noise is obtained by ${\bf\Sigma}^{(ite+1)}=\sigma^2_{(ite+1)}\textbf{I}_M$.

When the iteration converges, the parameters are denoted as ${\textit{\textbf{{H}}}}_{\rm EM}$ and ${\bf\Sigma}_{\rm EM}$. For the symbols engaged in the EM algorithm, the detection scheme based on the EM algorithm is different from that by the optimal ML algorithm. We have the following proposition.

\textit{Proposition 1}: The symbols in the EM algorithm can be detected by the following decision rule:
\begin{equation}\label{detect}
{\hat {\textbf{\textit{x}}}_{{\rm HEM},t}} = \textit{\textbf{s}}_{\mathop {\arg \max }\limits_{i} \hat{\tau}_{t,i}}
\end{equation}
where $\{\hat{\tau}_{t,i}\}$ is $\{{\tau}_{t,i}\}$ in E-step in the last iteration.

\textit{Proof}:
Note that $\hat\tau_{t,i}$ is the expectation of the hidden variable $z_{t,i}$ when the parameters ${\textit{\textbf{H}}}_{\rm EM}$ and ${\bf\Sigma}_{\rm EM}$ are given. It also infers the clustering probability of the transmitted symbol $\textit{\textbf{x}}_t$ being $\textit{\textbf{s}}_i$ in the codeword set $\mathcal{S}$. The symbol soft decisions are iteratively refined until the iteration converges, and the maximum value in $\{\hat{\tau}_{t,1}, \hat{\tau}_{t,2},... \hat{\tau}_{t,K}\}$ denotes the soft decision of $\textit{\textbf{x}}_t$.
Thus, we can directly detect the received symbols in the EM algorithm while estimating the channel by iteratively implementing E-step and M-step without extra calculations.

To obtain the detection results of the received signals in a frame from $\{\hat{\tau}_{t,i}\}$, all symbols need to engage into the iterative process of the EM algorithm, which will require a great deal of calculation operations. Different from the traditional LS estimation, the EM algorithm can fully extract the channel information with iterative progress so that the channel can be estimated accurately by only a few of the received symbols.
Thus, we use less number of the received symbols in the iteration of the EM algorithm to reduce the computation amount. The received symbols used in the EM algorithm for extracting the channel information are denoted as the sample symbols. The implement of the EM algorithm with the sample symbols can obtain the detection results of the sample symbols and the estimated channel matrix $\textit{\textbf{H}}_{\rm EM}$. The remaining received symbols in the frame can be detected by the ML detector using the estimated channel matrix $\textit{\textbf{H}}_{\rm EM}$.

It is well known that the EM algorithm is an alternative solution for the maximum likelihood estimation problem when the observed data are not complete. Inappropriate initialization will lead to the unsatisfactory local convergence.
We initialize the channel matrix by exploiting few pilots using the LS algorithm by (\ref{HLS}) and (\ref{SLS}). The inverse operation in (\ref{HLS}) requires that $\textit{\textbf{X}}_{pi}$ is a column full rank matrix. We set the length of the pilot sequence least as $L=N$.

%

\begin{algorithm}[t]

\caption{HEM-ML Algorithm for PMH signal detection}

\textbf{Inputs}: Received symbol matrix $\textit{\textbf{Y}}$, $\varepsilon$; \\
\textbf{Outputs}: ${\hat {\textbf{\textit{X}}}_{\rm HEM-ML}}$;
\begin{algorithmic}[1]

\STATE\textbf{Step a}: Initialization

\STATE Initialize the channel matrix with the first $N$ pilot symbols using the LS algorithm by (\ref{HLS}) and (\ref{SLS});

\STATE\textbf{Step b}: Detect the sample symbols and obtain CSI

\STATE Implement the EM algorithm with the next $R$ sample symbols;

\REPEAT

\STATE E-step: Calculate $\{{\tau}_{t,i}\}$ by (\ref{tau});

\STATE M-step: Update $\textit{\textbf{H}}^{\left( {ite{\rm{ + }}1} \right)}$ and $\sigma _{{\left( {ite + 1} \right)}}^2$ by (\ref{H}) and (\ref{sigma});

\UNTIL {$LL \left( {\textbf{\textit{Y}},\textbf{\textit{Z}}|{\theta}^{(ite+1)}}\right)-LL \left( {\textbf{\textit{Y}},\textbf{\textit{Z}}|{\theta}^{(ite)} } \right)<\varepsilon$}

\STATE ${\textit{\textbf{{H}}}}_{\rm EM}=\textit{\textbf{H}}^{\left( {ite{\rm{ + }}1} \right)}$, ${\bf\Sigma}_{\rm EM}={\bf\Sigma}^{\left( {ite{\rm{ + }}1} \right)}$, $\{\hat{\tau}_{t,i}\}=\{{\tau}^{(ite+1)}_{t,i}\}$;

\STATE Detect the $R$ sample symbols by (\ref{detect}) with $\{\hat{\tau}_{t,i}\}$;

\STATE\textbf{Step c}: Detect the rest of the symbols in the frame

\STATE Detect the remaining $T-(N+R)$ symbols using the estimated channel matrix ${\textit{\textbf{{H}}}}_{\rm EM}$ by (\ref{ML}).

\end{algorithmic}
\end{algorithm}

\begin{figure}[!hbt]
\renewcommand{\figurename}{Fig.}
\centering
\includegraphics[width=8.7cm]{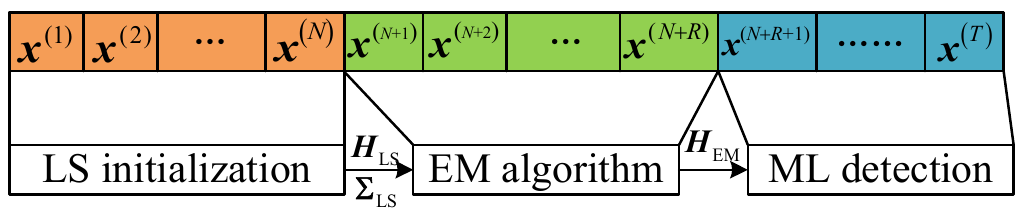}%
\label{receiver}
\vspace{-1em}
\caption{The frame structure of the HEM-ML detection scheme.}
\end{figure}

We choose the $R$ received symbols after the $L$ pilot symbols as the sample symbols. The complete HEM-ML detection scheme and the frame structure are described in \textbf{Algorithm 1} and Fig. 2. In HEM-ML, the detection process of the rest of the received symbols is still done in a brute-force way by the ML detector with the estimated $\textit{\textbf{H}}_{\rm EM}$.
Note that the number of constellation points $K$ increases exponentially with the number of antennas $N$ and the number of bits per antenna $k$.
In the next section, we propose a new detection algorithm for further reducing the computational complexity of the remaining $T-(L+R)$ symbols from exponential to linear with respect to the number of antennas without loss of detection precision as compared with traversal search in HEM-ML.


\section{Hypersphere clustering scheme based on EM Algorithm with KD-tree detection (HEM-KD)}
In this section, we introduce a signal detector design, HEM-KD with low complexity, based on KD-tree to detect the remaining $T-(L+R)$ PMH signals. First, the parameter information obtained from the $R$ sample symbols by the EM algorithm is used for building the spatial KD-tree codebook. Then, the signal detection problem is transformed into an NNS problem. Generally, the KD-tree and the NNS problem have been widely studied in the field of computer vision. KD-tree is a data structure that stores data in the $k$-dimensional space for fast searching. This inspires us to construct a spatial receiver codebook for the PMH signals in a $2N$ (real) dimensional space to speed up the detection process \cite{KD_tree}.
We next describe our proposed detection scheme HEM-KD and show that our detection scheme can achieve the same BER performance as the brute-force HEM-ML but with much less time complexity.


\subsection{Spatial Receiver KD-tree Codebook Construction}

First, we introduce the construction process of the spatial receiver codebook based on KD-tree with the channel information obtained from the sample symbols by EM algorithm. The KD-tree spatial codebook is a binary tree that makes use of each dimension of the constellation point to efficiently store the receiver codewords in the $2N$-dimensional real Euclidean space.

After being transferred through the channel $\textbf{\textit{H}}$, the constellation points of the transmitter codewords in the $2N$ dimensional real space are changed in amplitude and phase at the receiver. However, they can still be clustered on the hypersphere. In Section \uppercase\expandafter{\romannumeral3}, we can obtain the receiver codeword information $\{{\textbf{\textit{H}}}_{\rm EM}\textbf{\textit{s}}_i\}$ in (\ref{tau}) when the iteration converges. The receiver codebook is denoted as ${\cal S'}=\{\textbf{\textit{y}}'_1,\textbf{\textit{y}}'_2,
...,\textbf{\textit{y}}'_K\}$ where $\textbf{\textit{y}}'_i={\textbf{\textit{H}}}_{\rm EM}\textbf{\textit{s}}_i$ denotes the receiver codeword corresponding to the transmitter codeword $\textbf{\textit{s}}_i$ for $i\in \left\{ {1,...,K} \right\}$.

\begin{algorithm}[t]

\caption{KD-tree Codebook Construction Algorithm}

\textbf{Inputs}: Receiver codebook $\mathcal{S}'$; \\
\textbf{Outputs}: KD-tree codebook $\mathcal{K}$;\\
\textbf{Initialization}: $r=0, \cal W=\cal S'$;
\begin{algorithmic}[1]
\REPEAT

\IF {$|\cal W|$=1}

\STATE Mark the only point in the set $|\cal W|$, and not set left/right branch.

\ELSE

\STATE \textbf{Step 1}: Sort all points in $\cal W$ in ascending order at dimension $r$.
\STATE \textbf{Step 2}: Mark the point in the median at the $r$ dimension and record the current split dimension index $r$ of the marked point.

\STATE \textbf{Step 3}: Allocate the points in front of the marked points into set $\cal W_L$ and the points behind the marked points into set $\cal W_R$. $\cal W_L$ and $\cal W_R$ are the new data sets.

\STATE \textbf{Step 4}: The new data sets $\cal W_L$ and $\cal W_R$ are assigned as the left and right branches of the current marked node.

\STATE \textbf{Step 5}: $r\leftarrow(r+1)$mod$2M$.

\ENDIF

\UNTIL {every point in $\cal S'$ is marked}
\end{algorithmic}
\end{algorithm}

As mentioned in the system model, from the perspective of regarding the real part and imaginary part of the receiver codeword entries in different dimensions, the receiver codewords constellation ${\cal S'}$ is in a 2\textit{M}-dimensional real-valued coordinate system. We begin to build a KD-tree codebook to store those receiver codewords. The process of constructing a KD-tree is equivalent to continuously splitting the $2M$-dimensional space with a hyperplane paralleled to the coordinate axis to form a series of $2M$-dimensional rectangular regions.

We denote $\cal W$ as the data point set which will generate a marked node at each time. $\textit{r}\in\{0,...,2M-1\}$ is the split dimension index of the marked node to indicate the split being performed along the $r$ dimension. The split node is the median node of the points in the data point set in the $r$-th dimension to ensure the codebook being a balanced tree. If there are even points in the set, we choose the left or the right point next to the median point. The proposed codebook construction process is a recursive process and each iteration generates a marked point until every point in the codebook is marked. The details of the construction process are shown in \textbf{Algorithm 2}. $|\cal W|$ denotes the number of elements in $\cal W$.

After having obtained the KD-tree codebook $\mathcal{K}$, we detect the remaining received symbols $\textbf{\textit{y}}_t$ $(t=N+R+1,...,T)$ along the tree structured codebook. For every received symbol $\textbf{\textit{y}}_t$, the detection process is the same. Thus, for simplicity, we denote $\textbf{\textit{y}}_t$ as $\textbf{\textit{y}}$ in the sequential detection process. In the detection process, the search index is defined to track the search process. The node where the search index is located is defined as the current node, denoted as ${\rm {P}}$. ${\rm{P}}_{\rm r}$$=p$ denotes the value of the $r$-th dimension of the current node ${\rm {P}}$ being $p$. $\hat{{\textbf{\textit{x}}}}_{tem}$ denotes the temporary optimal node which represents the current optimal decision result. $d$ denotes the geometric distance between $\hat{{\textbf{\textit{x}}}}_{tem}$ and $\textbf{\textit{y}}$, $d'$ denotes the geometric distance between ${\rm {P}}$ and $\textbf{\textit{y}}$, and $l$ denotes the geometric distance between the split hyperplane and $\textbf{\textit{y}}$. The complete symbol detection process is summarized in \textbf{Algorithm 3}.
At the beginning, the search index is located at the root node. The split information refers to the split dimension index and the split hyperplane of the current node P.

\begin{figure}[!hbt]
\renewcommand{\figurename}{Fig.}
\centering
\includegraphics[height=4cm]{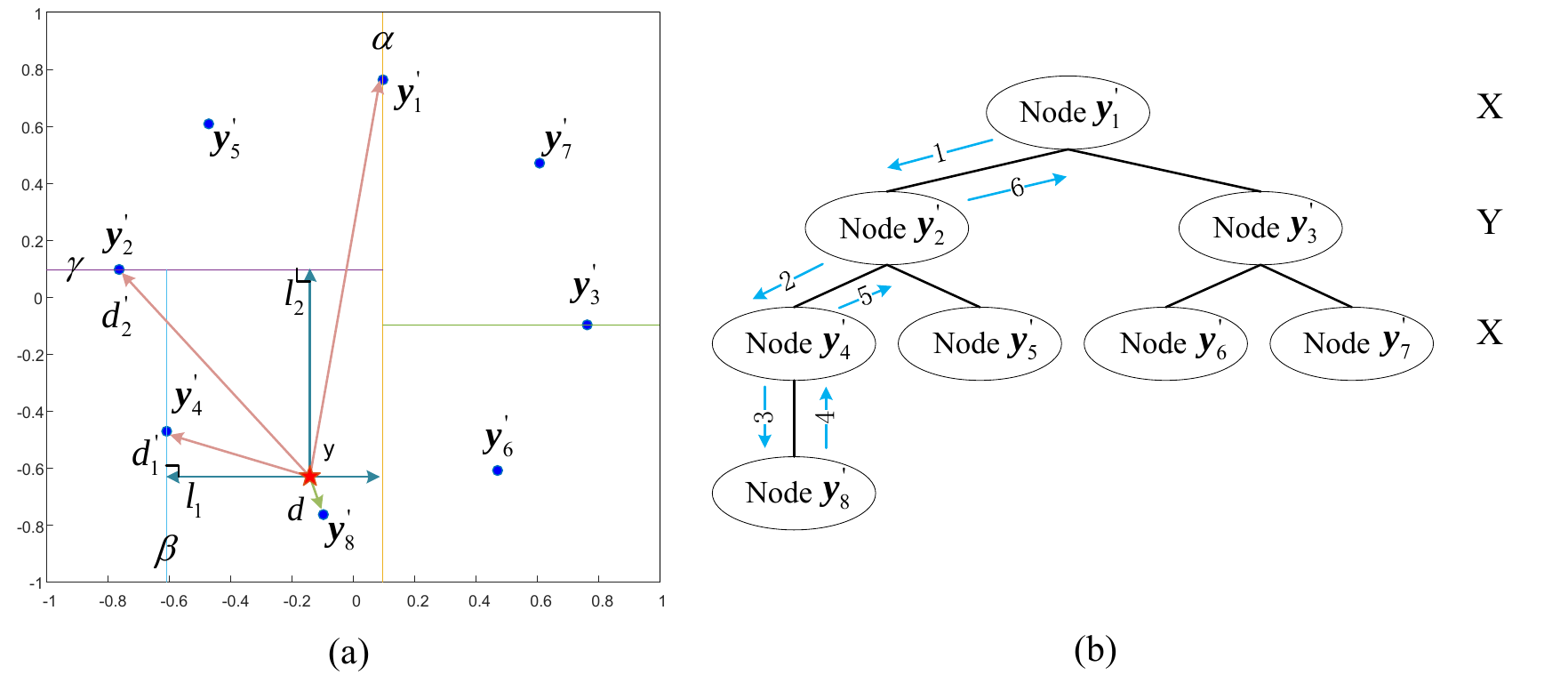}%
\label{receiver}
\vspace{-1em}
\caption{Two-dimensional KD-tree codebook and detection.}
\end{figure}
\vspace{-1em}

\begin{algorithm}[!htb]

\caption{Signal Detection Scheme Based on KD-tree Spatial Codebook}

\textbf{Inputs}: KD-tree codebook $\mathcal{K}$, received signal $\textbf{\textit{y}}$; \\
\textbf{Outputs}: detected signal ${\hat {\textbf{\textit{x}}}_{\rm HEM-KD}}$;

\begin{algorithmic}[1]

\STATE \textbf{Step a}:

\WHILE {(the search index doesn't reach a leaf)}

\STATE Read the split dimension index $r$ of P. $\rm{P}^r=\textit{p}$.

\IF  {($\textbf{\textit{y}}^r<p$)}

\STATE Move the search index to the left child node of P.

\ELSE

\STATE Move the search index to the right child node of P.

\ENDIF

\ENDWHILE

\STATE Mark P as a visited node.

\IF {($\hat{{\textbf{\textit{x}}}}_{tem}$ has existed)}

\IF {($d'<d$)}

\STATE $\hat{{\textbf{\textit{x}}}}_{tem}\leftarrow$P

\ENDIF

\ELSE

\STATE  $\hat{{\textbf{\textit{x}}}}_{tem}\leftarrow$P

\ENDIF

\STATE\textbf{Step b}:\\

\WHILE {(P is not the root node)}

\STATE \textbf{Step c}:\\

\STATE Move the search index to the parent node of P.

\IF {(P is not visited)}

\STATE  Mark P as a visited node.

\IF {$d'<d$}

\STATE $\hat{{\textbf{\textit{x}}}}_{tem}\leftarrow$P

\ENDIF


\IF {$l\geq d$}

\STATE Return to \textbf{Step b}.


\ELSE

\IF {(P has another child node)}

\STATE Move the search index to another child node

\STATE Return to \textbf{Step a}.

\ENDIF

\ENDIF

\ELSE

\STATE Return to \textbf{Step c}.

\ENDIF

\ENDWHILE


\STATE ${\hat {\textbf{\textit{x}}}_{\rm HEM-KD}}\leftarrow\hat{{\textbf{\textit{x}}}}_{tem}$

\end{algorithmic}
\end{algorithm}

To better understand the KD-tree codebook construction and the detection procedure, we explain the process through an example. We suppose that both the transmitter and the receiver have one antenna and the channel coefficiency is a complex scalar. The codewords are distributed on the two-dimensional real space ($x$ and $y$) and we set the number of bits per antenna to be 3, i.e., $k=3$. The size of the transmitter codebook is $K=8 $. The receiver codebook ${\cal S'}=\{\textbf{\textit{y}}'_1,\textbf{\textit{y}}'_2,
...,
\textbf{\textit{y}}'_8\}$ is respectively mapped to 8 transmitter codewords. The layout of the receiver codebook is shown in Fig. 3(a).

Now we construct the spatial KD-tree codebook. Split dimension index $r=0$ indicates that the first split is performed along the $x$ dimension and the split point is the median of the 8 points on the $x$ dimension. However, the size of the receiver codebook $|\cal W|$ is even. Here, we mark Node ${\textbf{\textit{y}}'_1}$ and record the current split dimension index $r=0$ as shown in Fig. 3(b). Node ${\textbf{\textit{y}}'_1}$ is noted as the first layer of the tree. Nodes $\{{\textbf{\textit{y}}'_2},{\textbf{\textit{y}}'_4},{\textbf{\textit{y}}'_5},{\textbf{\textit{y}}'_8}\}$ are allocated into set $\cal W_L$ of Node ${\textbf{\textit{y}}'_1}$ because their $x$ coordinates are smaller than that of Node ${\textbf{\textit{y}}'_1}$. Likewise, Nodes $\{{\textbf{\textit{y}}'_3},{\textbf{\textit{y}}'_6},{\textbf{\textit{y}}'_7}\}$ are allocated into set $\cal W_R$ of Node ${\textbf{\textit{y}}'_1}$. Then, $r\!=\!(0+1)\!\!\!\!\mod2=1$ represents the next split to be performed along the $y$ dimension in the new data sets $\cal W_L$ and $\cal W_R$. Nodes ${\textbf{\textit{y}}'_2}$ and ${\textbf{\textit{y}}'_3}$ are marked in the left and right branches of Node ${\textbf{\textit{y}}'_1}$, and they are noted as the second layer of the tree. Repeat this process until all nodes are marked. The three layer KD-tree receiver codebook and its spacial split are shown in Fig. 3(a) and Fig. 3(b).

Next, we illustrate the method to detect the received signal $\textbf{\textit{y}}$ based on the spatial KD-tree codebook. First, the search index is at the root node, i.e., Node ${\textbf{\textit{y}}'_1}$. The current node P is Node ${\textbf{\textit{y}}'_1}$. According to the split information of Node ${\textbf{\textit{y}}'_1}$, $r=0$, and the split hyperplane $\alpha$ referred to the split line in this example with two dimensions; $\textbf{\textit{y}}^0<{\textbf{\textit{y}}'_1}^0$ indicates that the $x$ coordinate of $\textbf{\textit{y}}$ is less than that of Node ${\textbf{\textit{y}}'_1}$. So, the search index moves towards left to Node ${\textbf{\textit{y}}'_2}$. Now the current node P refers to Node ${\textbf{\textit{y}}'_2}$ whose split dimension index is $r=1$ and $\textbf{\textit{y}}^1<{\textbf{\textit{y}}'_1}^1$.
Continue this procedure until the search index reaches Node ${\textbf{\textit{y}}'_8}$. P$\leftarrow{\textbf{\textit{y}}'_8}$. Then, we mark Node ${\textbf{\textit{y}}'_8}$ as a visited node. Yet, the temporarily chosen node $\hat{{\textbf{\textit{x}}}}_{tem}$ has not been defined. We denote Node ${\textbf{\textit{y}}'_8}$ as $\hat{{\textbf{\textit{x}}}}_{tem}$, implying that Node ${\textbf{\textit{y}}'_8}$ is the closest node to $\textbf{\textit{y}}$ so far. Since node ${\textbf{\textit{y}}'_8}$ is not the root node, we continue to implement Step c. The search index moves to the parent node of Node ${\textbf{\textit{y}}'_8}$, i.e., Node ${\textbf{\textit{y}}'_4}$.
The current node becomes Node ${\textbf{\textit{y}}'_4}$. Since Node ${\textbf{\textit{y}}'_4}$ has not been visited, we mark Node ${\textbf{\textit{y}}'_4}$ as a visited node. The distance between $\textbf{\textit{y}}$ and node ${\textbf{\textit{y}}'_4}$, $d_1'$, is more than that between $\textbf{\textit{y}}$ and $\hat{{\textbf{\textit{x}}}}_{tem}$(Node ${\textbf{\textit{y}}'_8}$), $d$, i.e., $d_1'>d$. $l_1$ is the distance between $\textbf{\textit{y}}$ and the split line $\beta$ of Node ${\textbf{\textit{y}}'_4}$. $l_1>d$ implies that no node in the other branch of Node ${\textbf{\textit{y}}'_4}$ is closer to $\textbf{\textit{y}}$ than $\hat{{\textbf{\textit{x}}}}_{tem}$.
Node ${\textbf{\textit{y}}'_4}$ is still not the root node, and thus we return to Step c. We move the search index to the parent node of Node ${\textbf{\textit{y}}'_4}$, i.e., Node ${\textbf{\textit{y}}'_2}$, which is the current node now. Mark Node ${\textbf{\textit{y}}'_2}$ as a visited node and compare the distance between Node ${\textbf{\textit{y}}'_2}$ and $\hat{{\textbf{\textit{x}}}}_{tem}$(Node ${\textbf{\textit{y}}'_8}$) to $\textbf{\textit{y}}$. We find Node ${\textbf{\textit{y}}'_8}$ is still the optimal detected node.
$l_2$ is the distance between $\textbf{\textit{y}}$ and the split line $\gamma$ of Node ${\textbf{\textit{y}}'_2}$, and $l_2>d$. Node ${\textbf{\textit{y}}'_2}$ is not the root node, and thus we move the search index to Node ${\textbf{\textit{y}}'_1}$. P is Node ${\textbf{\textit{y}}'_1}$ and we mark Node ${\textbf{\textit{y}}'_1}$ as a visited node. Likewise, we exclude the detection probabilities of Node ${\textbf{\textit{y}}'_3}$, ${\textbf{\textit{y}}'_6}$, ${\textbf{\textit{y}}'_7}$ to lower the detection time complexity. We finish the procedure and get  ${\hat {\textbf{\textit{x}}}_{\rm EM-KD}}\leftarrow\hat{{\textbf{\textit{x}}}}_{tem}$, i.e., ${\hat {\textbf{\textit{x}}}_{\rm EM-KD}}$ is Node ${\textbf{\textit{y}}'_8}$. Node ${\textbf{\textit{y}}'_8}$ is corresponding to the transmitter codeword $\textbf{\textit{s}}_8$. Thus, the detected result is $\textbf{\textit{s}}_8$.

In this section, we introduce a low-complexity symbol detection scheme HEM-KD for the remaining received symbols based on KD-tree algorithm with the constellation information obtained from the sample symbols by the EM algorithm. We emphasize that the detection result of the HEM-KD algorithm is the same as that of HEM-ML algorithm in Section \uppercase\expandafter{\romannumeral3}. This will be demonstrated by the simulation results in Section \uppercase\expandafter{\romannumeral6}. The complete detection process is summarized in \textbf{Algorithm 4} and Fig. 4.

\begin{algorithm}[H]

\caption{HEM-KD Algorithm for PMH signal detection}

\textbf{Inputs}: Received symbol matrix $\textit{\textbf{Y}}$, $\varepsilon$; \\
\textbf{Outputs}: ${\hat {\textbf{\textit{X}}}_{\rm HEM-KD}}$;
\begin{algorithmic}[1]

\STATE\textbf{Step a}: Initialization

\STATE Initialize the channel matrix with the first $N$ pilot symbols using the LS algorithm by (\ref{HLS}) and (\ref{SLS});

\STATE\textbf{Step b}: Detect the sample symbols and obtain the receiver codewords

\STATE Implement the EM algorithm for the next $R$ sample symbols;

\REPEAT

\STATE E-step: Calculate $\{{\tau}_{t,i}\}$ by (\ref{tau});

\STATE M-step: Update $\textit{\textbf{H}}^{\left( {ite{\rm{ + }}1} \right)}$ and $\sigma _{{\left( {ite + 1} \right)}}^2$ by (\ref{H}) and (\ref{sigma});

\UNTIL {$LL \left( {\textbf{\textit{Y}},\textbf{\textit{Z}}|{\theta}^{(ite+1)}}\right)-LL \left( {\textbf{\textit{Y}},\textbf{\textit{Z}}|{\theta}^{(ite)} } \right)<\varepsilon$}

\STATE $\{{\textit{\textbf{{H}}}}_{\rm EM}\textbf{\textit{s}}_i\}=\{\textit{\textbf{H}}^{\left( {ite{\rm{ + }}1} \right)}\textbf{\textit{s}}_i\}$, $\{\hat{\tau}_{t,i}\}=\{{\tau}^{(ite+1)}_{t,i}\}$;

\STATE Detect the $R$ sample symbols by (\ref{detect}) with $\{\hat{\tau}_{t,i}\}$;

\STATE\textbf{Step c}: Detect the rest of the symbols in the frame

\STATE Construct the KD-tree receiver codebook $\mathcal{K}$ by \textbf{Algorithm 2} with $\{{\textbf{\textit{H}}}_{\rm EM}\textbf{\textit{s}}_i\}$.

\STATE Detect the remaining $T-(N+R)$ symbols by \textbf{Algorithm 3}.

\end{algorithmic}
\end{algorithm}

\vspace{-2em}

\begin{figure}[!hbt]
\renewcommand{\figurename}{Fig.}
\centering
\includegraphics[width=8.8cm]{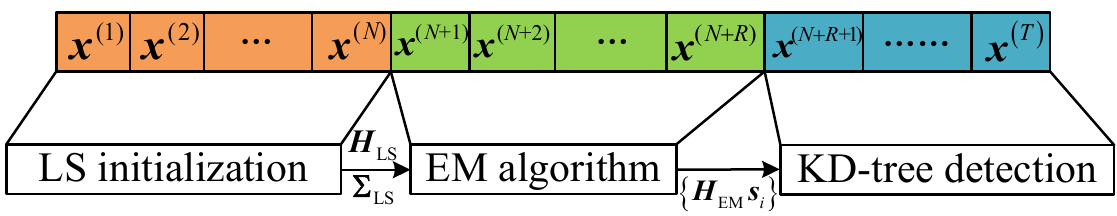}%
\label{receiver}
\vspace{-1em}
\caption{The frame structure of HEM-KD detection scheme.}
\end{figure}
\vspace{-3.5em}
\section{computational complexity analysis}

{In this section, we analyze the computational complexity of the introduced four detection schemes for comparison, namely, the optimal ML detection scheme, the LS detection scheme, the proposed HEM-ML detection scheme, and the proposed HEM-KD detection scheme.}
\vspace{-1em}
\subsection{Optimal ML Detection Scheme with Perfect CSI}

In the optimal ML detection scheme, with the known CSI, the computational complexity mainly comes from (\ref{ML}) with $\mathcal{O}(NK)$ of multiplications and $\mathcal{O}(K)$ of comparisons. For $T$ symbols in a frame, the total computational complexity for one frame is $\mathcal{O}(KT(N+1))$.
\vspace{-1em}
\subsection{Least Square Algorithm with Pilots Sequence}

{In the LS detection scheme, the computational complexity mainly comes from the channel matrix estimation (\ref{HLS}) and signal detection process (\ref{LS}). We assume that the length of the pilot sequence is $L$. The computational complexity of (\ref{HLS}) is $\mathcal{O}(NL^2+ML^2+L^3+MNL)$. The signal detection process (\ref{LS}) of the remaining $T-L$ symbols has the same computational complexity as the optimal ML detection scheme with perfect CSI, namely, $\mathcal{O}(NK)$ of multiplications and $\mathcal{O}(K)$ of comparisons for each symbol. For $T$ symbols and $L$ pilots sequences in a frame, the total computational complexity for one frame is $\mathcal{O}(L^3+L^2(N+M)+MNL+K(T-L)(N+1))$.}

\vspace{-1em}

\subsection{HEM-ML Scheme}


In the HEM-ML scheme, we analyze the computational complexity in the  three steps. In \textbf{step a}, the $N$ pilot symbols are used for the initialization by the LS algorithm. The computational complexity is $\mathcal{O}(2N^3+2MN^2)$. Then, the initial value is used in \textbf{step b} to perform the EM algorithm for the $R$ sample symbols. The computational complexity of the EM algorithm contains two steps, i.e., E-step and M-step.

\begin{itemize}
  \item \textbf{E-step}: The E-step is to update $\{{\tau}_{t,i}\}$ in (\ref{E}) for $t=N+1,...,N+R$ and $i=1,...,K$, and has $\mathcal{O}(KNR)$ multiplication operations.
  \item \textbf{M-step}: The M-step is to update $\textit{\textbf{H}}$ and $\sigma^2$ by (\ref{H}) and (\ref{sigma}) with $\mathcal{O}(KMN+2KNR)$ computational complexity.
\end{itemize}

For each sample symbol, the computational complexity of finding the maximum value of $\{\hat{\tau}_{t,i}\}$ $(i=1,...,K)$ in (\ref{detect}) is $\mathcal{O}(K)$. In \textbf{step c}, the detection of the remaining $T-(N+R)$ symbols is performed by the ML detector with $\mathcal{O}(K(N+1)(T-(N+R)))$ computational complexity.

For $I$ iterations in the EM algorithm, the total computational complexity for one frame is $\mathcal{O}(2N^3+2MN^2+KNI(M+3R)+K(N+1)(T-(N+R)))$.
\vspace{-1em}
\subsection{HEM-KD Scheme}

The HEM-KD scheme has the same computational complexity $\mathcal{O}(2N^3+2MN^2+KNI(M+3R))$ as the HEM-ML scheme in \textbf{step a} and \textbf{step b}.

In \textbf{step c}, we first analyze the computational complexity of constructing the spacial KD-tree codebook. Since the receiver codebook is a balanced KD-tree, the layer number of a KD-tree with $K$ codewords is $\log_2K-1$. In each layer, there is only one set having even number of elements. For the $n^{th}$ layer, there are $2^{n-1}-1$ sets with $\frac{K-2^{n-1}-2}{2^{n-1}}$ elements and one set with $\frac{K-2^{n-1}-2}{2^{n-1}}+1$ elements. Thus, the total computational complexity of constructing the KD-tree codebook is $\cal {O}$$((K+1)\log_2K-\frac{3}{2}K)$.

For the detection process for the remaining $T-(N+R)$ symbols, there is no need to calculate the Euclidean distances and find the maximum value in $K$ values like the optimal ML detection scheme and the HEM-ML scheme. The detection process is to organize the received point to the correct branch in the KD-tree. By using a balanced KD-tree, the average search complexity when the received symbols are uniformly distributed in the $2M$ dimensional real space is $\cal {O}$$(\log_2K)$. In practical communications, the received signals are likely distributed in a cluster of one codeword especially in high $E_b/N_0$ situation, and thus some branches of the KD-tree are excluded from the search. It significantly reduces the detection complexity to less than the average complexity $\cal {O}$$(\log_2K)$. At this step, we reduce the detection complexity from exponential $\cal {O}$$(2^{(kN)})$ to linear $\cal {O}$$(kN))$ with respect to the number of antennas $N$.
The total computational complexity for one frame is $\mathcal{O}(2N^3+2MN^2+(K+1)\log_2K-\frac{3}{2}K+KNI(M+3R)+ (\log_2K)(T-(N+R)))$.

From the computational complexity comparison of the four algorithms, we see that the two proposed detection schemes, HEM-ML and HEM-KD, may be more complex than the optimal ML detection scheme. This is attributed to the inverse operation in the initialization in (\ref{HLS}) and the iterative calculation to ensure convergence of the EM algorithm to the right global optimal point. The computational complexity of the optimal ML detection scheme is derived based on the know perfect CSI. As compared to the LS algorithm with a pilot sequence and the same pilot overhead, the proposed two detection schemes are shown to achieve better detection performance. The proposed HEM-KD scheme further reduces the computational complexity as compared to the proposed HEM-ML scheme in the detection process by reducing the search complexity from the exponential level to the linear level.

\section{Simulation Results}

In this section, simulation results are presented to show the detection performance of the proposed detection schemes HEM-ML and HEM-KD in comparison with the optimal ML detector and the LS detector for point-to-point MIMO systems with PMH modulation. Block flat-fading channel models are considered in the simulations. We assume that the receiver has the same number of antennas with the transmitter, namely, $M=N$, and there are 5000 symbols in each frame, i.e., $T=5000$. The threshold of the iteration in the EM algorithm is set to be $\varepsilon=10^{-6}$. $10^8$ Monte-Carlo runs are carried out to achieve reliable results. In the simulation, we consider the following two scenarios:

\begin{enumerate}
  \item [$\bullet$]\textit{Rayleigh Channel}
\end{enumerate}

For the Rayleigh channel, we assume that the entries of the channel matrix $\textbf{\textit{H}}_{\rm Ray}$ are jointly \textit{i.i.d.} and proper complex Gaussian random variables of variance $\sigma^2=1$.
\begin{enumerate}
  \item [$\bullet$]\textit{Millimeter Wave Channel}
\end{enumerate}

In the millimeter wave band, the propagation environment between the transmitter and the receiver is modeled as a geometric channel with ${N_{ray}}$ paths \cite{channel}. Furthermore, we assume the uniform linear array antenna configuration. For such an environment, the channel matrix can be written as
$\textbf{\textit{H}}_{\rm mmW}{\rm{ = }}\sqrt {\frac{{NM}}{N_{ray}}} \sum\limits_{{\rm{l}} = 1}^{N_{ray}} {{\alpha ^l}{\textbf{\textit{a}}_r}\left( {\phi _r^l} \right){\textbf{\textit{a}}_t}{{\left( {\phi _t^l} \right)}^H}}$,
where $\alpha ^l\sim \cal{CN}$$(0,1)$ is the complex gain of the $l$-th path between the transmitter and the receiver, and ${\phi _r^l}\in[0,2\pi)$ and ${\phi _t^l}\in[0,2\pi)$ are the angle of arrive (AoA) and angle of departure (AoD), respectively. Furthermore, ${\textbf{\textit{a}}_r}(\cdot)$ and ${\textbf{\textit{a}}_t}(\cdot)$ are the antenna array response vectors at the receiver and the transmitter, respectively. In a uniform linear array configuration with $N$ antenna elements, we have
${{\textbf{\textit{a}}}}\left( \phi  \right) = \frac{1}{{\sqrt N }}{\left[ {1,{e^{j\tilde k\tilde d\sin (\phi )}},...,{e^{j\tilde k\tilde d(N - 1)\sin (\phi )}}} \right]^T}$,
where $\tilde k=\frac{2\pi}{{\lambda }}$, $\lambda$ is the wavelength and $\tilde d=\frac{\lambda}{2}$ is the antenna spacing.
\vspace{-1em}
\subsection{Bit Error Rate Performance}

Fig. 5 shows the bit error rate (BER) performance versus bit energy to noise density (${E_{\rm b}}$/N0) for different numbers of bits per antenna $k$ with the fixed number of antennas $N=2$. We set that $R=100$ sample symbols are used for the EM algorithm in a frame in Figs. 5-7. We compare the BER performance of the optimal ML detector, the LS detector, HEM-ML, HEM-KD with the LS initialization and HEM-KD with random initialization.

We see that HEM-KD with random initialization leads to a poor performance because the iteration is prone to converge to a local optimal point especially when the initial canstellation points are randomly distributed in a multidemensional space. Furthermore, with the radius of the hypersphere of the PMH symbols $N$ fixed, the curves with $k=1$ perform better than the curves with $k=2$ because more bits per antenna $k$ enable more information to be carried by each PMH symbol. More transmitter codewords $K=Nk$ distributed on the surface of the hypersphere ${\mho_N}\left( N \right)$ implies shorter distance between the constellation points, which leads to a worse BER performance as shown in Fig. 5.

\begin{figure}[htbp]
\begin{minipage}[t]{0.32\linewidth}
\centering
\includegraphics[width=4cm]{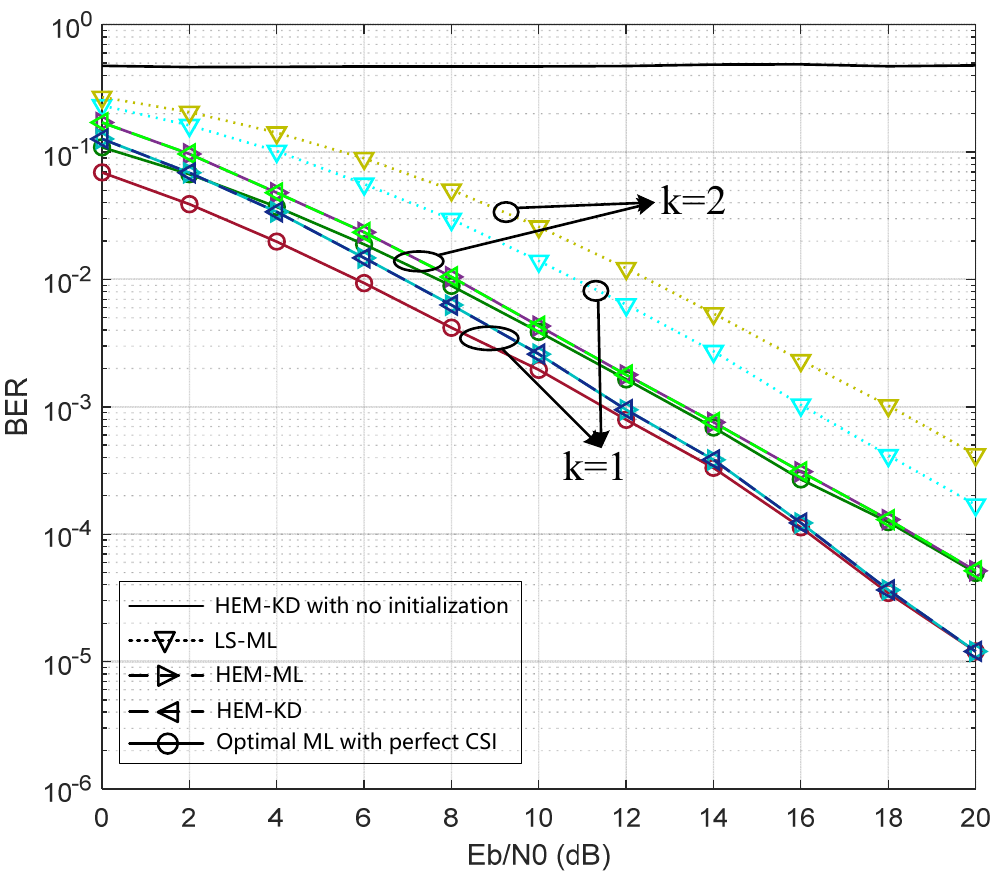}
\vspace{-1em}
\caption{BER performance versus $E_{\rm{b}}/N_{\rm{0}}$, $N=2$.}
\end{minipage}%
\begin{minipage}[t]{0.32\linewidth}
\centering
\includegraphics[width=4cm]{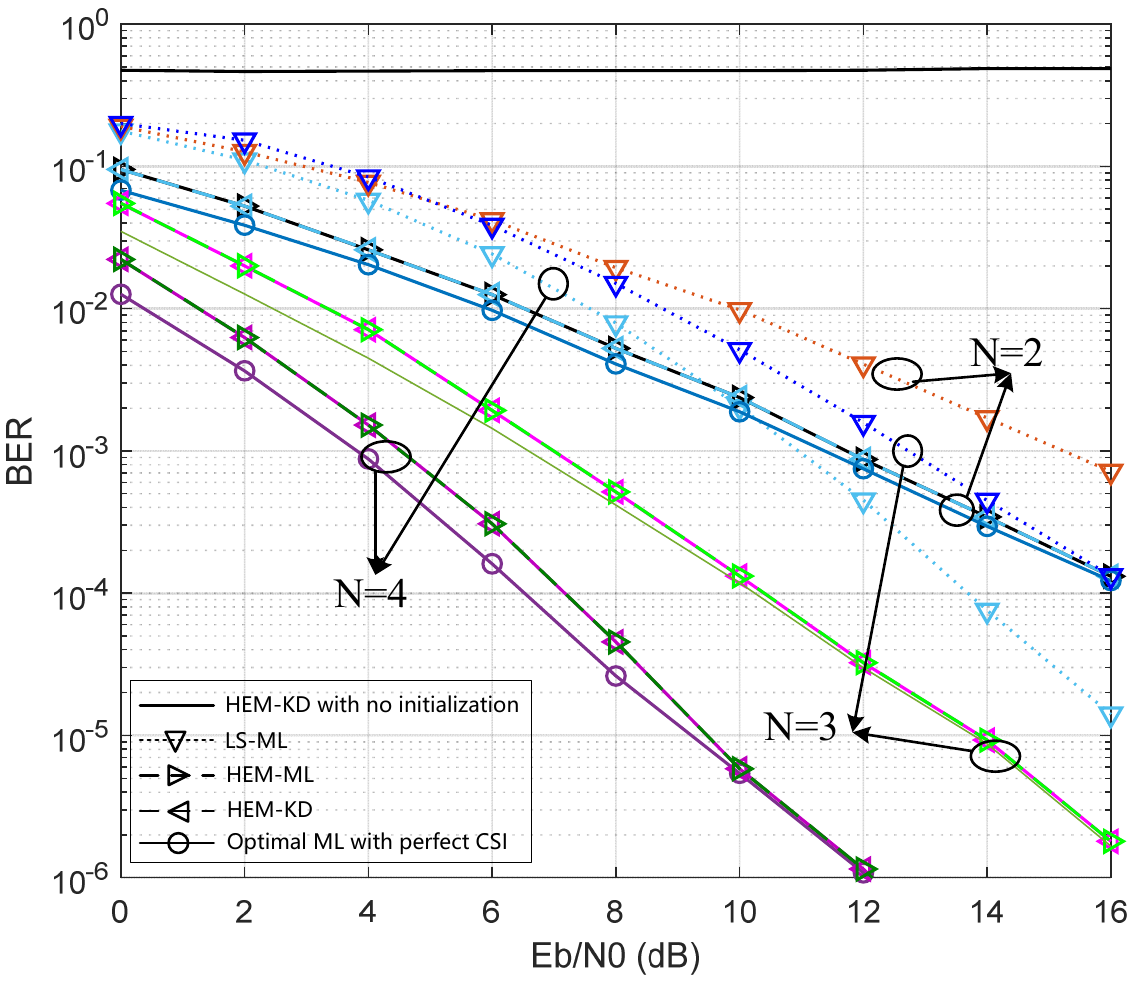}
\vspace{-1em}
\caption{BER performance versus $E_{\rm{b}}/N0$, $k=1$.}
\end{minipage}
\begin{minipage}[t]{0.32\linewidth}
\centering
\includegraphics[width=4cm]{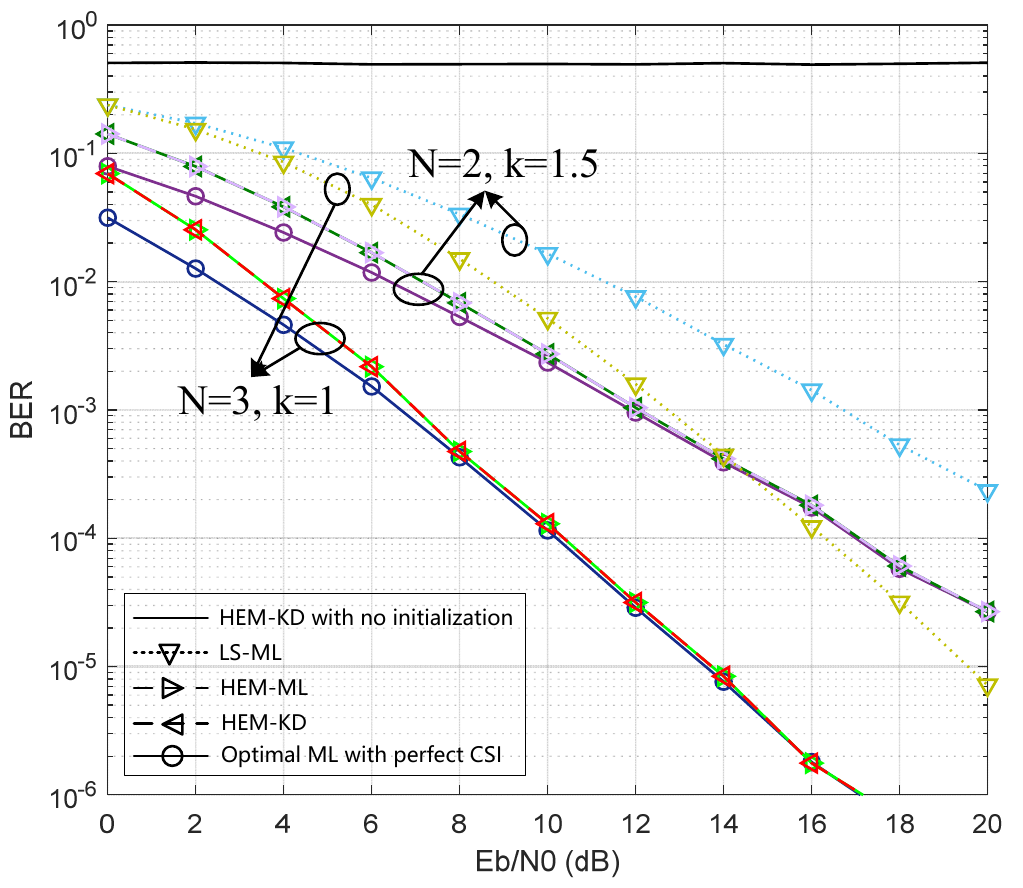}
\vspace{-1em}
\caption{BER performance versus $E_{\rm{b}}/N{0}$, $K=3$.}
\end{minipage}
\end{figure}


HEM-ML and HEM-KD perform better than the LS detection scheme because the EM algorithm can make full use of the sample symbols to extract the channel information as long as it has a good initialization. Compared with the LS algorithm which depends on the pilot symbols, the EM algorithm is able to achieve more accurate perameter estimation. The curves of HEM-ML/HEM-KD and ML are getting close as $E_{\rm b}/N0$ increases because the received symbols are more likely gathered around the cluster centroids for the high $E_{\rm b}/N0$ case. Besides, it is worth noting that the simulation results of HEM-ML and HEM-KD are exactly the same because the signal detection process along the KD-tree in the HEM-KD can obtain the same ``clostest" optimal point as the ML detection in the HEM-ML.

Fig. 6 shows the BER performance versus $E_{\rm b}/N0$ with a fixed number of bits per antenna $k=1$. It is observed that by increasing the number of antennas $N$, the BER performance increases as justified in \cite{PMH} that the minimum distance between the canstellation points grows with the number of antennas for a fixed number of bits per antenna.


In Fig. 7, we present the BER performance for different $N$ and $k$ settings with a fixed bit rate per symbol $K=kN=3$. The information carried by each symbol in each case is the same, implying that the number of the canstellation points is fixed. More antennas $N$ indicates the larger radius of the hypersphere ${\mho_N}\left( N \right)$ and hence the distance between the canstellation points grows, which leads to a better BER performance.
\vspace{-2em}

\subsection{Computational complexity}

Tabel \ref{tabel} shows the comparisons of the performance and computational complexity of the algorithms. $I_{\rm ave}$ denotes the average number of iterations in the EM algorithm. The number of the computation operations of one frame is bolded. We set $N=3$, $k=1$ and $E_{\rm b}/N0=10$dB. From the table, we can see that the EM algorithm has a higher efficiency in extracting the channel information than the LS algorithm. Only few received symbols used in the EM algorithm can lead to a precise estimation. Furthermore, as compared with HEM-ML, HEM-KD can greatly reduce the computational complexity without any BER performance loss.

\begin{figure}[htbp]
\begin{minipage}[t]{0.48\linewidth}
\centering
\includegraphics[width=4cm]{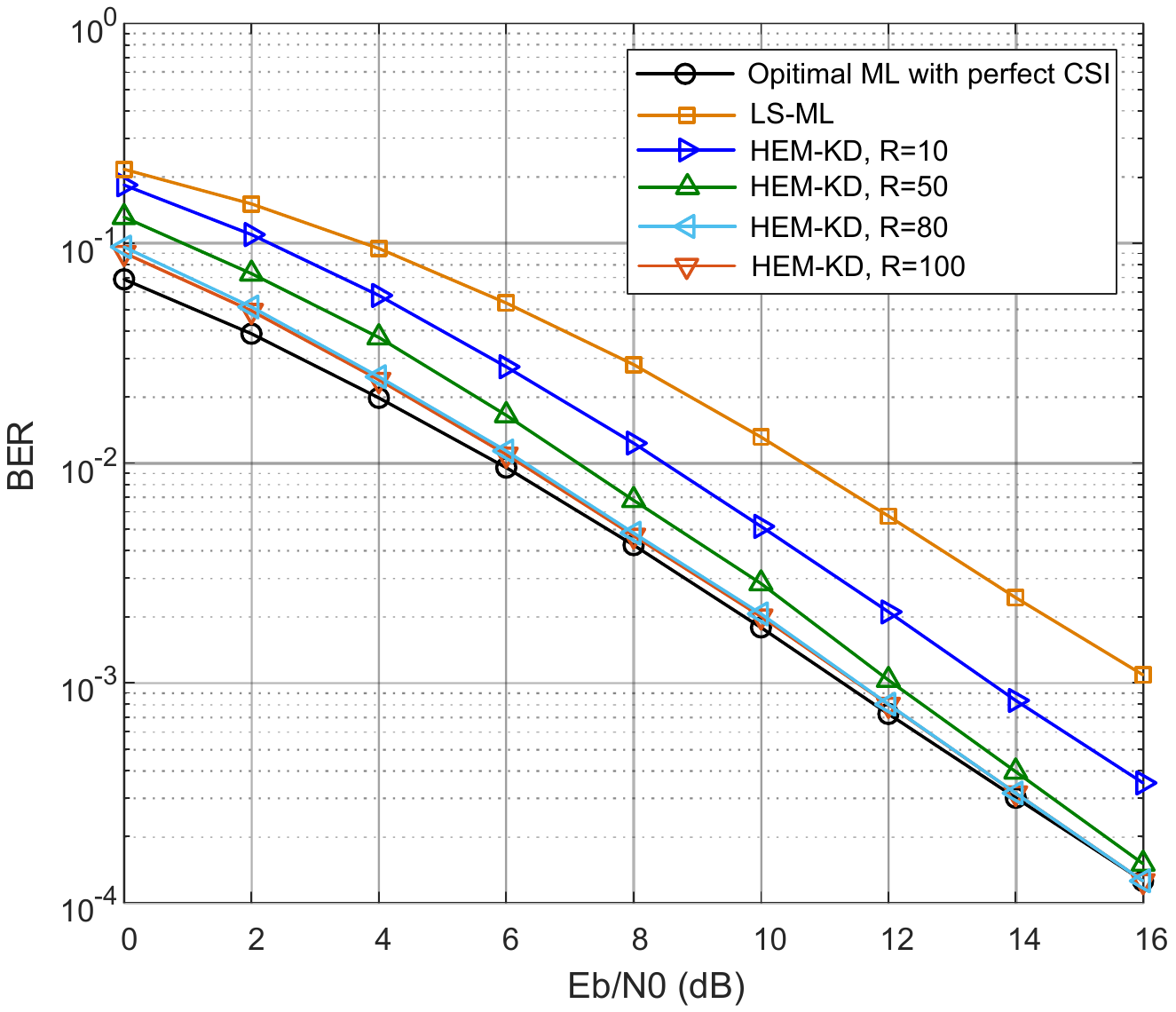}
\vspace{-1em}
\caption{BER performance versus $E_{\rm{b}}/N0$ of HEM-KD, $N=2$, $k=2$.}
\end{minipage}
\begin{minipage}[t]{0.48\linewidth}
\centering
\includegraphics[width=4cm]{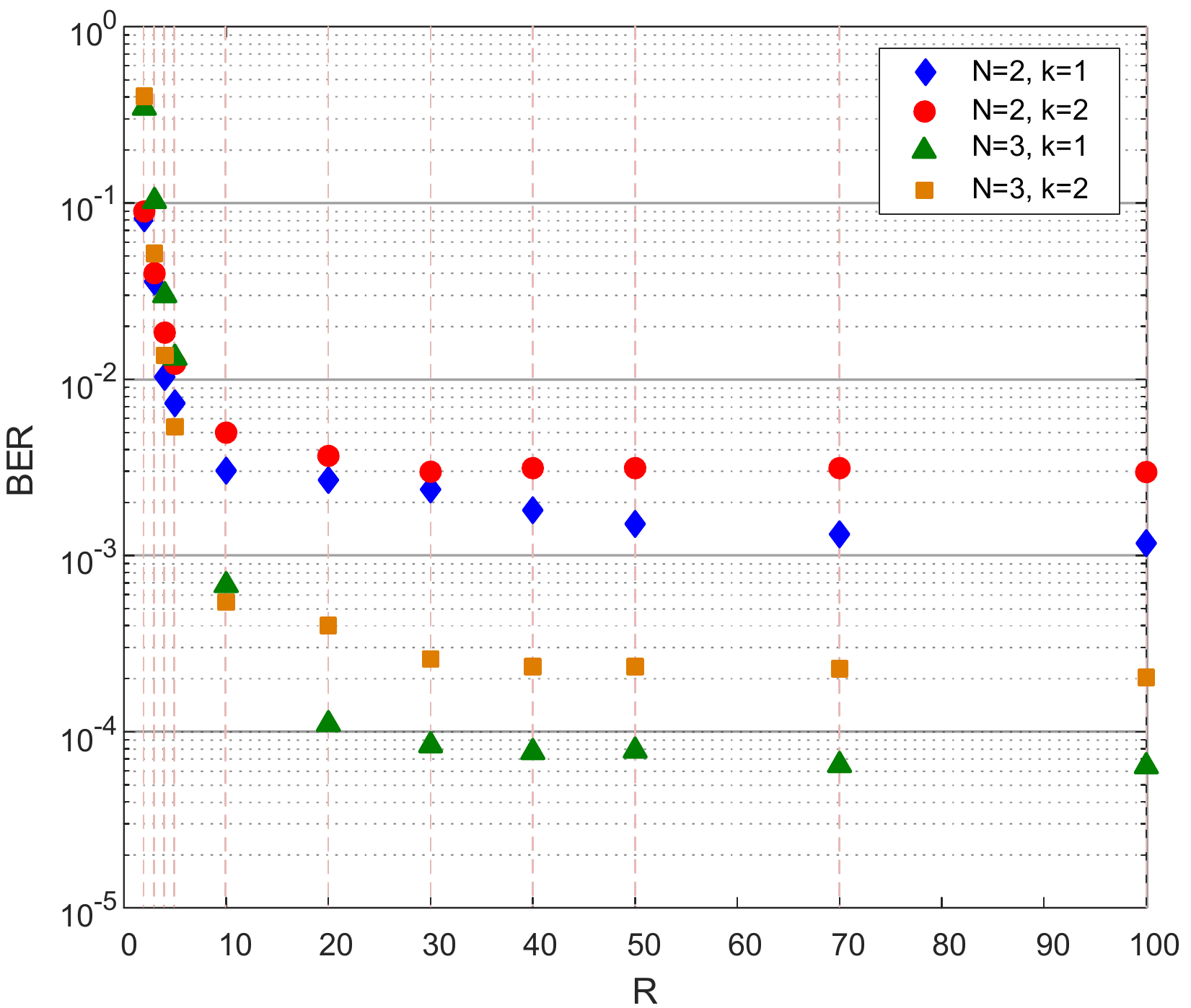}
\vspace{-1em}
\caption{BER performance versus $R$ in HEM-KD, $E_{\rm{b}}/N0=11$dB.}
\end{minipage}
\end{figure}

\renewcommand{\arraystretch}{1} 
\begin{table}[tp]
  \centering
  \fontsize{9}{8}\selectfont
  \begin{threeparttable}
  \caption{Performance and Computational Complexity Comparison}
  \label{tabel}
    \begin{tabular}{ccccccc}
    \toprule
    \multirow{2}{*}{Algorithm}&
    \multicolumn{2}{c}{Frame Parameter}&\multicolumn{3}{c}{Evaluation}\cr
    \cmidrule(lr){2-3} \cmidrule(lr){4-6}
    &$L$&$R$&Computation&$I_{\rm ave}$&BER\cr
    \midrule
    ML&--&--& \textbf{160000} &  -- &$1.15\times 10^{-4}$\cr
    \multirow{3}{*}{LS}&3&--&\textbf{160012}&--&$1.9\times 10^{-3}$\cr
    &60&--&\textbf{396220}&--&$2.2\times 10^{-4}$\cr
    &300&--&\textbf{27693100}&--&$1.18\times 10^{-4}$\cr
    \multirow{3}{*}{HEM-ML}&3&5&\textbf{161200}&3.13&$1.00\times 10^{-3}$\cr
    &3&20&\textbf{164200}&3.19&$1.97\times 10^{-4}$\cr
    &3&5000&\textbf{1253100}&3.48&$1.25\times 10^{-4}$\cr
    \multirow{3}{*}{HEM-KD}&3&5&\textbf{16451}&3.13&$1.00\times 10^{-3}$\cr
    &3&20&\textbf{19877}&3.19&$1.97\times 10^{-4}$\cr
    &3&5000&\textbf{1253100}&3.48&$1.25\times 10^{-4}$\cr
    \bottomrule
    \end{tabular}
    \end{threeparttable}
\end{table}

\vspace{-4em}
\subsection{Convergence Evaluation}


Fig. 8 shows the trend of the BER performance versus $E_{\rm{b}}/N0$ for different numbers of sample symbols $R$. We can see from the simulation result that the CSI can be better estimated with more $R$. The EM algorithm can perform better channel estimation as compared with the LS algorithm especially in high $E_{\rm{b}}/N0$ scenarios. As $R$ increases, the BER performance grades slowly to a limit.


Fig. 9 illustrates the BER performance under different numbers of received symbols used in the EM algorithm. We set $E_{\rm{b}}/N0=11$dB. We can see that the EM algorithm can extract the channel information with a small number of received symbols. The longer distance between the codewords leads to the requirement of more received symbols engaged in the EM algorithm.

Fig. 10 shows the normalized mean squared error (NMSE) ${\omega _\textbf{H}}$ of the estimated channel parameter in the EM algorithm versus the iteration counter. The number of bits per antenna is set to be $k=1$. The NMSE of the estimated channel is defined as ${\omega _\textbf{H}}{\rm{ = }}{{{{\left\| {{\textbf{H}_{\rm{EM}}}{\rm{ - }}\textbf{H}} \right\|}^2}} \mathord{\left/
{\vphantom {{{{\left\| {{\textbf{H}_{\rm{EM}}}{\rm{ - }}\textbf{H}} \right\|}^2}} {{{\left\| \textbf{H} \right\|}^2}}}} \right.
\kern-\nulldelimiterspace} {{{\left\| \textbf{H} \right\|}^2}}}$. We can see that the iteration converges faster when $N=3$ than $N=2$. For different numbers of sample symbols in a frame $R=100$ and $R=200$, the convergence rates have negligible differences. In addition, it is noted that $E_{\rm b}/N0$ effects the estimation accuracy. When we set $N=2$ and $R=100$, the curve with $E_{\rm b}/N0=15\rm{dB}$ tends to flatten after the 14-${th}$ iteration at $10^{-5}$ while the curve with $E_{\rm b}/N0=20\rm{dB}$ converges with a smaller NMSE.

Fig. 11 presents the number of iterations of the EM estimation versus $E_{\rm b}/N0$. We set the curve with $N=2,k=1,R=100$ as the benchmark. From the line with $N=2,k=2,R=100$, we see that with the fixed radius of the hypersphere $(N)$, a larger $k$ means more codewords, thus, the distances between the points become smaller. More codewords and less distance require more iterations to cluster the symbols. The curve with $N=3,k=1,R=100$ shows that although there are more codewords, the distances between the constellation points become longer, which is consistent with the conclusion shown in Fig. 6. On the whole, the distance between the constellation points is the key factor to the speed of the convergence. Besides, the curve with $N=2,k=1,R=200$ shows that the convergence process slightly improves as the number of samples increases because more samples carry more information of the channel.


\begin{figure}[htbp]
\begin{minipage}[t]{0.32\linewidth}
\centering
\includegraphics[width=4cm]{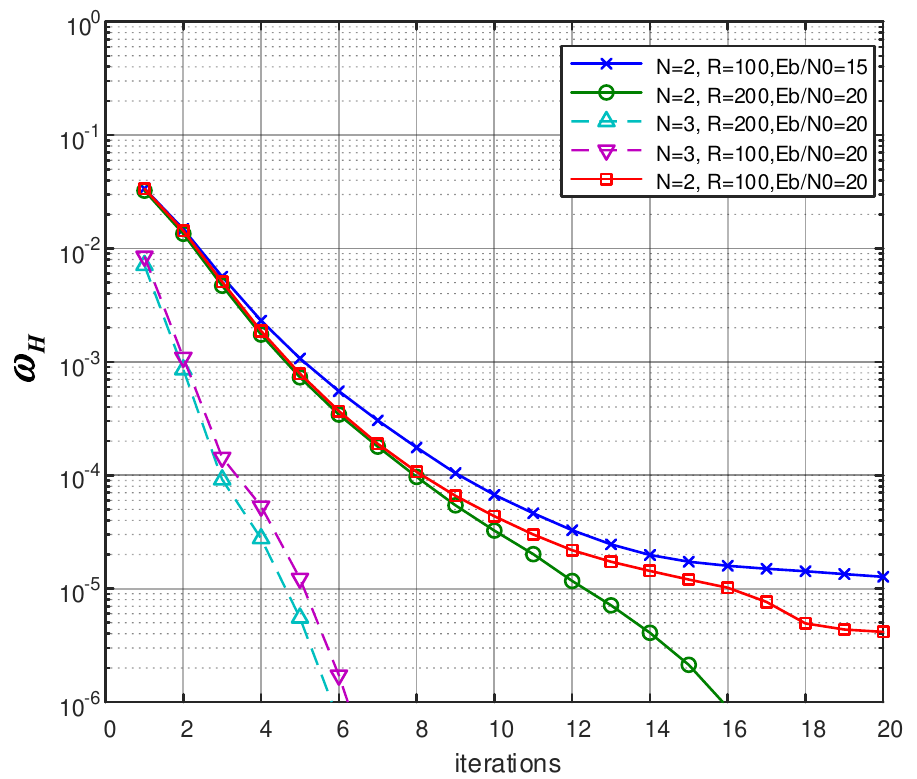}
\vspace{-1em}
\caption{NMSE versus the number of iterations of HEM-KD, $k=1$.}
\end{minipage}
\begin{minipage}[t]{0.32\linewidth}
\centering
\includegraphics[width=4cm]{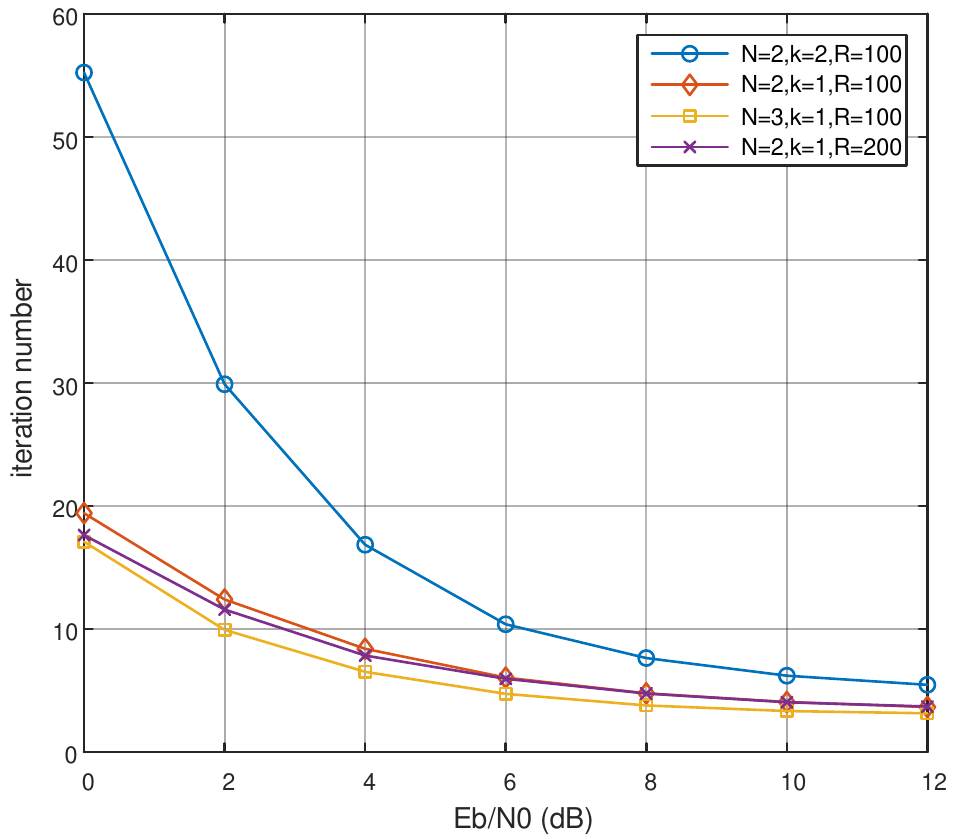}
\vspace{-1em}
\caption{Total number of iterations versus $E_{\rm{b}}/N0$ of HEM-KD.}
\end{minipage}
\begin{minipage}[t]{0.32\linewidth}
\centering
\includegraphics[width=4cm]{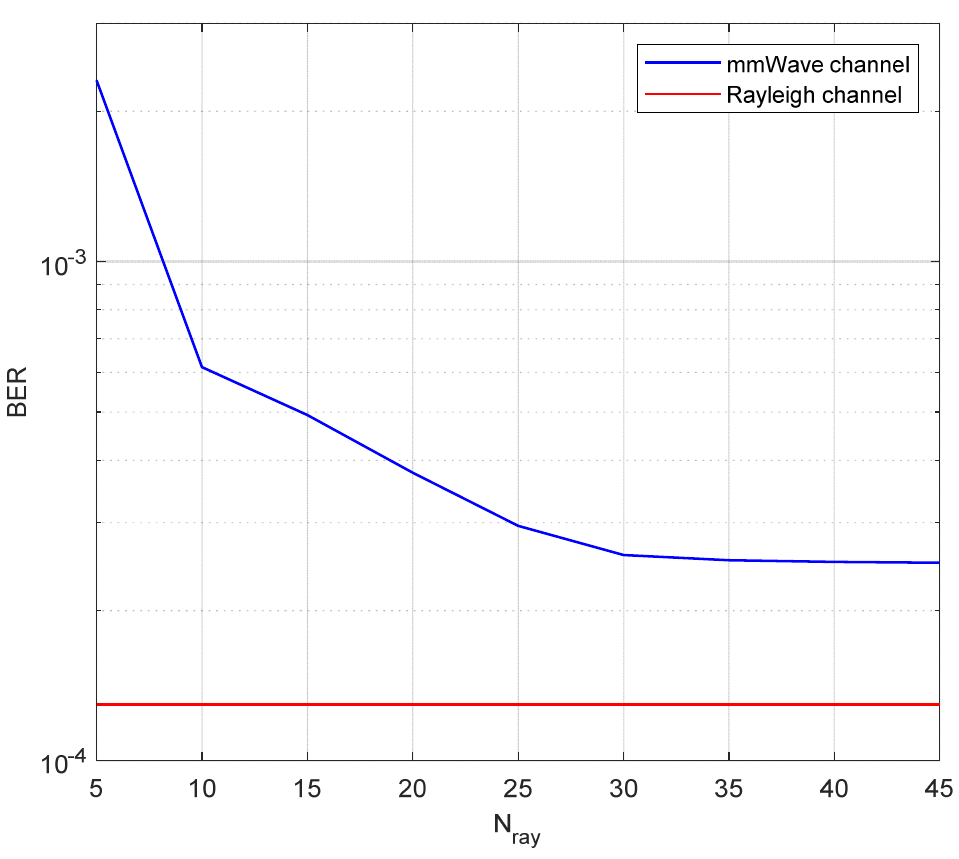}
\vspace{-1em}
\caption{BER performance of HEM-KD in mmWave channel and Rayleigh channel $N=3$, $k=1$, $R=100$, $ {E_{\rm b}/N0}=10$.}
\end{minipage}
\end{figure}

%
%
%

\vspace{-3em}
\subsection{Channel Evaluation}

{Fig. 12 shows the BER performance of the proposed HEM-KD detection scheme in the mmWave channel versus the number of paths of the mmWave channel $N_{ray}$. The parameters are set as $N=3$, $k=1$ and ${E_{\rm b}/N0}=10$. Using the performance of HEM-KD in the Rayleigh channel as a benchmark, we find that HEM-KD shows a better detection performance in the Rayleigh channel than in the mmWave channel because the mmWave channel is prone to be sparse and only few angles are available to transfer the information if the path is rare. This propagation feature of the mmWave channel affects the spatial integrity of the multi-dimensional transmitted PMH signals, as verified in the figure that the BER curve of HEM-KD in the mmWave channel gets closer to that in the Rayleigh channel with richer paths $N_{ray}$.}
\vspace{-1em}
\section{Conclusions}

In this paper, we proposed two machine learning-based signal detection schemes for PMH signals in $\textit{load-modulated}$ MIMO systems: HEM-ML and HEM-KD. The EM algorithm can efficiently extract the information carried by the signals than the traditional pilot-driven method. HEM-ML exploits the information carried in a few of the received symbols to jointly estimate the channel parameters and update the soft decision of this part of the received symbols iteratively. In HEM-KD, a data structure KD-tree in machine learning is employed to further reduce the detection complexity by constructing the spatial receiver codebook based on KD-tree and detecting the symbols along the KD-tree codebook. The results show that the proposed two detection schemes, by leveraging machine learning, have achieved great error-rate performance as compared to the optimal ML detector and the LS estimator, especially in a high SNR scenario.

\vspace{-1em}

\end{document}